\documentstyle[twocolumn,aps]{revtex}

\begin{document}
\draft
\title{Magnetic field dependence of the exciton energy in a quantum disk}
\author{K. L. Janssens, F. M. Peeters$^{\circ }$ and V. A. Schweigert$^{\dagger }$}
\address{Departement Natuurkunde, Universiteit Antwerpen (UIA),\\
Universiteitsplein 1, B-2610 Antwerpen, Belgium}
\date{\today}
\maketitle

\begin{abstract}
The groundstate energy and binding energy of an exciton, confined in a
quantum disk, are calculated as a function of an external magnetic field.
The confinement potential is a hard wall of finite height. The diamagnetic
shift is investigated for magnetic fields up to 40$T$. Our results are
applied to $In_{y}Al_{1-y}As/Al_{x}Ga_{1-x}As$ self-assembled quantum dots
and very good agreement with experiments is obtained if the light hole is
assumed to be involved in the exciton formation. Furthermore, we
investigated the influence of the dot size on the diamagnetic shift by
changing the disk radius. The exciton excited states are found as a function
of the magnetic field. The relative angular momentum is not a good quantum
number and its value changes with the magnetic field strength.
\end{abstract}


\section{Introduction}

Recently, there has been much interest in the study of quantum dots, which
are structures in which the charge carriers are confined in all three
dimensions. Especially the self-assembled quantum dots \cite{bimboek} are
considered to be very promising for possible applications, such as quantum
dot lasers \cite{grundmrs}, due to their large confinement energy and high
optical quality. The dots are formed by the Stranski-Krastanow growth mode
in which a material, e.g. $In_{y}Al_{1-y}As$, is deposited on another
material with a substantially different lattice parameter, e.g. $%
Al_{x}Ga_{1-x}As$ \cite{petroff}$.$ The lattice mismatch, which is required
for this growth process, is typically about $4\%$ \cite{pryor}$.$ Initially,
the growth is two dimensional, but after a critical thickness of a few
monolayers, coherent islands are formed due to strain effects. The shape of
the formed islands is not well known, but is expected to resemble a lens or
a pyramid. The density, size and shape of the dots are strongly dependent on
the growth conditions. Typical sizes of dots vary between the basis size of
7 to 20$nm$ and a height of a few nanometers. The density of the dots is of
the order 10$^{11}cm^{-2}$ \cite{grundmrs}.

The properties of confined excitons have been the subject of many
theoretical studies. Bryant \cite{bryant} used variational and
configuration-interaction representations to study excitons in quantum
boxes. Later, matrix diagonalization techniques were used to study the
exciton energy in a quantum dot with parabolic confinement potential. Song 
{\it et al. }\cite{song} studied the effect of non circular symmetric
structures, and Halonen {\it et al.} \cite{halonen}\ studied the influence
of a magnetic field. More recently, Pereyra {\it et al.} \cite{ulloa}\
investigated magnetic field and quantum confinement asymmetry effects on
excitons, again for the case of parabolic confinement. These studies have
shown a strong competition between the quantum dot size, Coulomb interaction
and magnetic confinement.

In the present work, we approximate the quantum dots by a quantum disk with
a hard-wall confinement of finite height \cite{goff}, as found in
self-assembled quantum dots, and include the mass mismatch between the dot
and barrier material. We present a theoretical study of the effect of an
external magnetic field on the properties of an exciton in the quantum disk,
fully taking into account the Coulomb interaction between the electron and
the hole. The groundstate energy and binding energy of the exciton are
studied as a function of the magnetic field. This allows us to determine the
diamagnetic shift of the exciton, which we find in very good agreement with
the experimentally observed shift by Wang {\it et al.} \cite{wang}. In most
of the previous theoretical work , this diamagnetic shift was only
determined for very low values of the magnetic field, where the confinement
energy is larger than the cyclotron energy, and could be approximated by $%
e^{2}\left\langle \rho ^{2}\right\rangle B^{2}/8\mu $ \cite
{wang,bayer,polimeni}. In our calculations, we consider magnetic fields up
to 40$T.$ For such large magnetic fields, the weak field approximation is no
longer valid, because now the cyclotron energy overcomes the confinement
energy and the particles will act rather as free particles in a magnetic
field \cite{nomura}. We find that the magnetic field dependence of the
diamagnetic shift can be very closely approximated by $\beta B^{2}/(1+\alpha
B)$. To be able to make a valid comparison between theory and experiment, we
considered for our simulations $In_{0.55}Al_{0.45}As$ quantum dots, which
were experimentally studied by Wang {\it et al.} \cite{wang}.

The paper is organized as follows. In Sec. II, we present the theoretical
model and explain our method of solution. The results for the exciton
groundstate and the comparison with the experimental results of Ref.~\cite
{wang} are presented in Sec. III. In Sec. IV, we describe the effect of
changing the disk radius on the exciton energy and diamagnetic shift. The
results for the exciton energy spectrum are presented in Sec. V. Our results
are summarized in Sec. VI.

\section{Theoretical model}

The Hamiltonian describing our system is given by 
\begin{equation}
H=%
\mathrel{\mathop{\stackrel{2}{\sum }}\limits_{j=1}}%
H_{j}({\bf r}_{j})+V_{c}({\bf r}_{1}-{\bf r}_{2})\text{ ,}  \label{eq1}
\end{equation}
with 
\begin{equation}
H_{j}=({\bf p}_{j}-\frac{q_{j}}{c}{\bf A}_{j})\frac{1}{2m_{j}({\bf r})}({\bf %
p}_{j}-\frac{q_{j}}{c}{\bf A}_{j})+V_{j}({\bf r}_{j})\text{ ,}  \label{eq2}
\end{equation}
where the indices $j=1,2$ correspond to the electron and the hole with
masses $m_{1},m_{2}$, respectively, $V_{j}(\rho _{j},z_{j})=0(\rho
_{j}<R,\left| z_{j}\right| <d/2),V_{j,o}$ (otherwise) is the confinement
potential with $R$ the radius of the quantum disk and $d$ its thickness, $%
\rho _{j}=\sqrt{x_{j}^{2}+y_{j}^{2}},$ $V_{c}({\bf r})=-e^{2}/\epsilon
\left| {\bf r}\right| ,$ and $q_{j}=\mp e.$ Here and below the upper and
lower sign correspond to electron and hole, respectively. For convenience we
will sometimes also use the notations $e,h$ instead of $1,2.$ We allow for a
difference in mass between the dot region and the region outside the dot: $%
m_{j}({\bf r})=m_{w,j}$ inside the disk and $m_{j}({\bf r})=m_{b,j}$ outside
the disk. In our numerical work, we used the following values for the
physical parameters: $\epsilon =12.71,$ $m_{w,e}=0.076m_{0},$ $%
m_{b,e}=0.097m_{0}$, $m_{w,h}=m_{b,h}=0.45m_{0}$, $V_{e,o}=258meV,$ and $%
V_{h,o}=172meV,$ which are typical for the $%
In_{0.55}Al_{0.45}As/Al_{0.35}Ga_{0.65}As$ system.

Using cylindrical coordinates ${\bf r}_{j}=(z_{j},\rho _{j},\phi _{j})$ the
one-particle Hamiltonian takes the form 
\begin{eqnarray}
H_{j} &=&-\frac{\text{$\hbar $}^{2}}{2}\left( \frac{\partial }{\partial z_{j}%
}\frac{1}{m_{j}}\frac{\partial }{\partial z_{j}}+\frac{1}{\rho _{j}}\frac{%
\partial }{\partial \rho _{j}}\frac{\rho _{j}}{m_{j}}\frac{\partial }{%
\partial \rho _{j}}+\frac{1}{\rho _{j}^{2}m_{j}}\frac{\partial ^{2}}{%
\partial \phi _{j}^{2}}\right)  \nonumber  \label{eq3} \\
&&\mp \frac{i}{2}\text{$\hbar $}\omega _{c,j}\frac{\partial }{\partial \phi
_{j}}+\frac{1}{8}m_{j}\omega _{c,j}^{2}\rho _{j}^{2}+V_{j}(z_{j},\rho _{j})%
\text{ ,}
\end{eqnarray}
where $\omega _{c,j}=eB/m_{j}c$ are the electron and hole cyclotron
frequencies and the vector potential is taken in the symmetrical gauge ${\bf %
A=}\frac{1}{2}B\rho {\bf e}_{\phi }.$

The one-particle wave functions are separable $\Psi _{j}(z,\rho ,\phi )=(1/%
\sqrt{2\pi })e^{il\phi }\xi _{j,i}^{l}(z_{j},\rho _{j}),$ where $l=0,\pm
1,\pm 2,...$ is the angular momentum, and the wave functions $\xi
_{j,i}^{l}(z_{j},\rho _{j})$ are eigenfunctions of the Hamiltonian 
\begin{eqnarray}
H_{j}^{l} &=&-\frac{\text{$\hbar $}^{2}}{2}\left( \frac{\partial }{\partial
z_{j}}\frac{1}{m_{j}}\frac{\partial }{\partial z_{j}}+\frac{1}{\rho _{j}}%
\frac{\partial }{\partial \rho _{j}}\frac{\rho _{j}}{m_{j}}\frac{\partial }{%
\partial \rho _{j}}\right) +\frac{\text{$\hbar $}^{2}l^{2}}{2m_{j}\rho
_{j}^{2}}  \nonumber  \label{eq4} \\
&&\pm \frac{l}{2}\text{$\hbar $}\omega _{c,j}+\frac{1}{8}m_{j}\omega
_{c,j}^{2}\rho _{j}^{2}+V_{j}(z_{j},\rho _{j})\text{ },
\end{eqnarray}
where the index $i$ denotes the eigenenergies of $H_{j}^{l}.$ As a
consequence of the axial symmetry of our problem, there is no coupling
between the wave functions with different values of the total angular
momentum $L.$ Therefore, we can construct the exciton wave function $\Psi
_{L}$ with fixed total momentum $L$ as the linear combination 
\begin{equation}
\Psi _{L}({\bf r}_{1},{\bf r}_{2})=%
\mathrel{\mathop{\stackrel{l_{m}}{\sum }}\limits_{l=-l_{m}}}%
\psi ^{l}({\bf \chi })e^{i\frac{l}{2}(\phi _{1}-\phi _{2})+i\frac{L}{2}(\phi
_{1}+\phi _{2})}\text{ ,}  \label{eq5}
\end{equation}
where the functions $\psi ^{l}({\bf \chi })$ obey the Schr\"{o}dinger
equation 
\begin{equation}
\mathrel{\mathop{\stackrel{2}{\sum }}\limits_{j=1}}%
H_{j}^{l}\psi ^{l}({\bf \chi })+%
\mathrel{\mathop{\stackrel{l_{m}}{\sum }}\limits_{l^{\prime }=-l_{m}}}%
V_{c}^{l-l^{\prime }}({\bf \chi })\psi ^{l^{\prime }}({\bf \chi })=E\psi
^{l}({\bf \chi })\text{ ,}  \label{eq6}
\end{equation}
with $E$ the eigenenergy, for brevity ${\bf \chi }$ denotes the coordinates $%
(z_{1},z_{2},\rho _{1,}\rho _{2}),$ $V_{c}^{l}$ is the matrix element of the
Coulomb interaction 
\begin{eqnarray}
V_{c}^{l}({\bf \chi }) &=&-\frac{e^{2}}{\epsilon }\frac{1}{2\pi }  \nonumber
\\
&&\times 
\mathrel{\mathop{\stackrel{2\pi }{\int }}\limits_{0}}%
d\phi \frac{e^{-il\phi }}{\sqrt{(z_{1}-z_{2})^{2}+\rho _{1}^{2}+\rho
_{2}^{2}-2\rho _{1}\rho _{2}\cos (\phi )}}\text{ },
\end{eqnarray}
$L_{m}=2l_{m}+1$ is the total number of angular harmonics in the expansion.

A common technique to solve the eigenvalue problem is to use an expansion of
the wave function in a suitable set of basis functions. For the typical
sizes of the quantum disks considered here, the exciton binding energy is
much smaller than the confinement energy. As a consequence, a natural choice
is to take the eigenfunctions $\xi _{j,i}^{l}$ of the one-particle
Hamiltonian. But for our present problem such an approach runs into
obstacles because of the enormous number of basis functions which are
required to obtain the binding energy with sufficient accuracy. Indeed,
using the one particle eigenfunctions for different values of the angular
momentum $l$ and quantum number $i=1,...I,$ one has to calculate $L_{m}I^{4}$
matrix elements of the Hamiltonian. In the present case of hard wall
confinement, the one dimensional eigenfunctions are too complicated in order
to obtain an analytical expression for the Coulomb matrix elements.
Therefore, a numerical integration procedure has to be used on the space
grid with size $N_{g}=(K\times N)^{2},$ where $K,$ $N$ are the numbers of
grid points for the longitudinal and transverse directions, respectively. In
principle, the difficulties in the calculation of the Coulomb matrix
elements can be avoided by applying an appropriate basis, for instance the
nonorthogonal Gaussian basis, which is widely employed in quantum chemical
simulations. But in this case there is an increase of the number of
functions, which are needed, leading to difficulties with diagonalizing a
large $L_{m}I^{2}\times L_{m}I^{2}$ non-sparse matrix. Note that for an
arbitrary basis, the number $I=i_{z}\times i_{r}$ is determined by the
number of one-particle wave functions in the longitudinal ($i_{z}$) and the
radial ($i_{r}$) directions. The total number of operations depends
crucially on the considered number of subbands $i_{z}$ in the $z$-direction.
For a small ratio $d/R$ of the longitudinal to transverse size of the
quantum disk as given before, we can limit ourself by taking only one
subband \cite{peeters,lamouche}.

\subsection{3D exciton problem}

For arbitrary values of the ratio $d/R$ we present a numerical technique
based on the use of a finite difference scheme. Let $z_{k},$ $(k=1,...,K),$ $%
\rho _{n},$ $(n=1,...,N)$ be some nonuniform space grid in the longitudinal
and transverse directions for both electron and hole coordinates. Using the
appropriate symmetry conditions for the ground wave function in the
longitudinal direction $\partial \psi /\partial z_{j}(z_{j}$=$0)=0$ we can
limit ourselves to the region $z_{j}>0.$ Thus, the first point of the $z$%
-grid corresponds to $z=0.$ The upper $(z_{K}>d/2)$ and right $(\rho _{N}>R)$
boundaries of the simulation region correspond to the barrier region where
the wave function and its derivatives go to zero. Therefore, the Neumann
conditions $\partial \psi /\partial z_{j}=0,$ $\partial \psi /\partial \rho
_{j}=0$ are employed for these boundaries. To obtain the finite difference
scheme for the one-particle Hamiltonian, including the discontinuous
behavior of the particles mass and external potential, we integrate the
expression over the square $(z_{k-1/2}<z<z_{k+1/2},\rho _{n-1/2}<\rho <\rho
_{n+1/2}),$ where the subgrids with noninteger indexes are determined by the
relations $z_{k+1/2}=(z_{k+1}+z_{k})/2,\rho _{n+1/2}=(\rho _{n+1}+\rho
_{n})/2,z_{-1/2}=\rho _{-1/2}=0.$ Substituting the finite difference
expressions for the derivatives of the wave function $\partial \psi
/\partial z(z=z_{k+1/2})=(\psi _{k+1}-\psi _{k})/(z_{k+1}-z_{k}),\partial
\psi /\partial \rho (\rho =\rho _{n+1/2})=(\psi _{n+1}-\psi _{n})/(\rho
_{n+1}-\rho _{n})$ we obtain the following finite difference scheme for the
one-particle Hamiltonian 
\begin{eqnarray}
(\hat{H}_{j}^{l}\psi )_{k,n} &=&-a_{j}^{k,n}\psi _{k+1,n}-c_{j}^{k,n}\psi
_{k-1,n}-b_{j}^{k,n}\psi _{k,n+1}  \nonumber  \label{eq8} \\
&&-d_{j}^{k,n}\psi _{k,n-1}+p_{j}^{k,n}\psi _{k,n}\text{ ,}
\end{eqnarray}
with the coefficients 
\begin{eqnarray}
a_{j}^{k\neq 1,n} &=&\text{$\hbar $}%
^{2}(1/m_{jz}^{k,n}+1/m_{jz}^{k-1,n})/2(z_{k}-z_{k-1})h_{z,k},  \nonumber \\
a_{j}^{k=1,n} &=&0,  \eqnum{9a} \\
c_{j}^{k\neq K,n} &=&\text{$\hbar $}%
^{2}(1/m_{jz}^{k,n}+1/m_{jz}^{k+1,n})/2(z_{k+1}-z_{k})h_{z,k},  \nonumber \\
c_{j}^{k=K,n} &=&0,  \eqnum{9b} \\
b_{j}^{k,n\neq 1} &=&\rho _{n-1/2}\text{$\hbar $}^{2}(1/m_{j\rho
}^{k,n}+1/m_{j\rho }^{k,n-1})/2(\rho _{n}-\rho _{n-1})h_{\rho ,n},  \nonumber
\\
b_{j}^{k,n=1} &=&0,  \eqnum{9c} \\
d_{j}^{k,n\neq N} &=&\rho _{n+1/2}\text{$\hbar $}^{2}(1/m_{j\rho
}^{k,n}+1/m_{j\rho }^{k,n+1})/2(\rho _{n+1}-\rho _{n})h_{\rho ,n},  \nonumber
\\
d_{j}^{k,n=N} &=&0,  \eqnum{9d}
\end{eqnarray}
\begin{eqnarray}
p_{j}^{k,n} &=&a_{j}^{k,n}+b_{j}^{k,n}+c_{j}^{k,n}+d_{j}^{k,n}+\frac{\text{$%
\hbar $}^{2}l^{2}}{2\rho _{n}^{2}m_{j}^{k,n}}\pm \frac{l}{2}\text{$\hbar $}%
\omega _{c,j}^{k,n}  \nonumber \\
&&+\frac{1}{8}m_{j}^{k,n}(\omega _{c,j}^{k,n})^{2}\rho _{n}^{2}+V_{j}^{k,n},
\eqnum{9e}
\end{eqnarray}
where $h_{z,k}=z_{k+1/2}-z_{k-1/2},$ $h_{\rho ,n}=(\rho _{n+1/2}^{2}-\rho
_{n-1/2}^{2})/2.$ Due to the discontinuity of the mass and the external
potential at the disk boundary, special care must be taken in the choice of
the expression for its grid values. In the expressions the averaged value of
the masses $m_{jz}^{k,n},m_{j\rho }^{k,n},m_{j}^{k,n}$ and potential $%
V_{j}^{k,n}$ are determined by the following relations $%
(m_{jz}^{k,n})^{-1}=h_{\rho ,n}^{-1}\int\nolimits_{\rho _{n-1/2}}^{\rho
_{n+1/2}}\rho m_{j}^{-1}(z$=$z_{k},\rho )d\rho ,$ $(m_{j\rho
}^{k,n})^{-1}=h_{z,n}^{-1}\int\nolimits_{z_{n-1/2}}^{z_{n+1/2}}m_{j}^{-1}(z,%
\rho $=$\rho _{n})dz,$ $(m_{j}^{k,n})^{-1}=h_{z,k}^{-1}h_{\rho
,n}^{-1}\int_{z_{k-1/2}}^{z_{k+1/2}}dz\int_{\rho _{n-1/2}}^{\rho
_{n+1/2}}\rho m_{j}^{-1}(z,\rho )d\rho ,$ $V_{j}^{k,n}=h_{z,k}^{-1}h_{\rho
,n}^{-1}\int_{z_{k-1/2}}^{z_{k+1/2}}dz\int_{\rho _{n-1/2}}^{\rho
_{n+1/2}}V_{j}(z,\rho )\rho d\rho .$

Once a finite difference Hamiltonian $\hat{H}=\delta _{l,l^{\prime
}}\sum_{j=1}^{2}\hat{H}_{j}^{l}+V_{c}^{l,l^{\prime }}\delta _{M,M^{\prime }}$
has been constructed, we have to develop a technique to obtain the ground
state of the sparse matrix $\hat{H}.$ Here $\delta _{i,j}$ is the unit
matrix, index $M$ denotes all indexes corresponding to the space grid. Note
that the number of non zero elements of the matrix $\hat{H}$ is only
proportional to $L_{m}^{2}N_{g}.$ This is a key distinction from the
commonly accepted expansion over basis functions, where this number
increases as the second power with the number of functions. However, the
size of our matrix is still large and therefore direct diagonalization
methods are not suitable for solving our problem. The best suitable approach
to find only the lowest eigenvalue $E_{g}$ and eigenvector $\Psi $ is the
inverse iteration method, where the eigenvector $\Psi ^{i}$ at the $i^{th}$
stage of the iteration is obtained by solving the following equation 
\begin{equation}
(\hat{H}-\lambda \delta _{l,l^{\prime }}\delta _{M,M^{\prime }})\overline{%
\Psi }^{i}=\Psi ^{i-1}\text{ ,}  \eqnum{10}  \label{eq14}
\end{equation}
with the subsequent normalization 
\begin{equation}
\Psi ^{i}=\overline{\Psi }^{i}/\sqrt{\left\langle \overline{\Psi }^{i},%
\overline{\Psi }^{i}\right\rangle }\text{ ,}  \eqnum{11}  \label{eq15}
\end{equation}
where the brackets $\left\langle \text{ },\text{ }\right\rangle $ stand for
scalar multiplication. The eigenenergy is obtained in the usual way $%
E_{g}^{i}=\left\langle \Psi ^{i},\hat{H}\Psi ^{i}\right\rangle .$ The value
of the parameter $\lambda <E_{g}$ is chosen such that a minimum absolute
value of the matrix $(\hat{H}-\lambda \delta _{l,l^{\prime }}\delta
_{M,M^{\prime }})$ corresponds to the ground state of the matrix $\hat{H}$.
There exist many numerical relaxation techniques to solve the boundary value
problem. Using standard methods one has to solve the equation with good
precision at each stage of the inverse iteration procedure. Here, we propose
a new technique, which generalizes in fact the commonly accepted
Gauss-Seidel methods \cite{numrec} with inverse iterations. The value of the
eigenvector $\Psi ^{i}$ for the mesh points $(l,m=k_{1},n_{1},k_{2},n_{2})$
is obtained by using the following relation 
\begin{equation}
\Psi _{i}=(\Psi _{i-1}+\alpha _{i-1}\Theta _{1}+\Theta _{2})/%
\mathrel{\mathop{\stackrel{2}{(\sum p_{j}+V_{c}^{ll}-\lambda )\text{ ,}}}\limits_{j=1}}%
\eqnum{12}  \label{eq16}
\end{equation}
where 
\begin{eqnarray}
\Theta _{1} &=&\sum\nolimits_{j=1}^{2}(c_{j}\Psi _{i-1}^{k_{j}+1}+d_{j}\Psi
_{i-1}^{n_{j}+1})  \nonumber \\
&&-\sum\nolimits_{l^{\prime }>l}^{l_{m}}V_{c}^{l,l^{\prime }}\Psi
_{i-1}^{l^{\prime }},  \eqnum{13a} \\
\Theta _{2} &=&\sum\nolimits_{j=1}^{2}(a_{j}\Psi _{i}^{k_{j}-1}+b_{j}\Psi
_{i}^{n_{j}-1})  \nonumber \\
&&-\sum\nolimits_{l=l_{m}}^{l<l^{\prime }}V_{c}^{l,l^{\prime }}\Psi
_{i}^{l^{\prime }}.  \eqnum{13b}
\end{eqnarray}
For the ground state $\overline{\Psi }_{i}=\Psi _{i-1}/(E_{g}-\lambda ),$ we
found that the maximum rate of convergency is realized by using the
following values of the parameters $\alpha _{i\neq 1}=1/(E_{g}^{i}-\lambda )$
and$\ \alpha _{i=1}=1.$

\subsection{2D exciton problem}

For quantum disks with large radius $R\gg d$ we use the adiabatic approach,
a technique which was already successfully applied in Refs.~\cite
{peeters,peeters2}. Within this approach, we can write the wavefunction as 
\begin{equation}
\psi _{l}({\bf \chi })=\psi _{1}(z_{1})\psi _{2}(z_{2})\psi _{l}(\rho
_{1},\rho _{2}),  \eqnum{14}  \label{eq19}
\end{equation}
where $\psi _{j}(z_{j})$ corresponds to the groundstate of the longitudinal
Hamiltonian 
\begin{equation}
H_{z,j}=-\frac{\hbar ^{2}}{2}\frac{\partial }{\partial z_{j}}\frac{1}{m_{j}}%
\frac{\partial }{\partial z_{j}}+V_{j,z}(z_{j}),  \eqnum{15}  \label{eq20}
\end{equation}
for electron $(j$=$1)$ and hole $(j$=$2),$ respectively. Since the wave
function penetrates only slightly into the barrier region in the radial
direction, the longitudinal behavior of the effective masses $m_{j}$ and the
confinement potentials $V_{j,z}$ can, to high accuracy, be approximated by $%
m_{j,z}=m_{j}(z,\rho _{j}=0),$ $V_{j,z}=V_{j}(z,\rho _{j}=0).$ Then the wave
function of the ground state has a simple form inside, $\psi _{j}(\left|
z\right| <d/2)=\cos (k_{j}z),$ and outside, $\psi _{j}(\left| z\right|
>d/2)=\exp (-\kappa _{j}\left| z\right| ),$ the disk, where $k_{j}=\sqrt{%
2m_{w,j}E_{0,zj}}/\hbar $ and $\kappa _{j}=\sqrt{2m_{b,j}(V_{o,j}-E_{0,zj})}%
/\hbar .$ The energy of the groundstate $E_{0,zj}$ is obtained from the
continuity of the wave function and conservation of the current $%
m^{-1}\partial \psi /\partial z$ at the boundary $(\left| z\right| =d).$
Substituting expression~(\ref{eq19}) into the Schr\"{o}dinger equation and
integrating out the $z_{j}$ coordinates by taking the average $\left\langle
\psi _{1}(z_{1})\psi _{2}(z_{2})\left| H\right| \psi _{1}(z_{1})\psi
_{2}(z_{2})\right\rangle =H^{2D},$ we obtain the effective two dimensional
Hamiltonian 
\begin{eqnarray}
H^{2D} &=&\sum_{j=1}^{2}\left[ \left( {\bf p}_{j}-\frac{q_{j}}{c}{\bf A}%
_{j}\right) \frac{1}{2m_{j}^{\prime }({\bf \rho }_{j})}\left( {\bf p}_{j}-%
\frac{q_{j}}{c}{\bf A}_{j}\right) +V_{j}^{\prime }({\bf \rho }_{j})\right] 
\nonumber \\
&&+V_{c}^{\prime }({\bf \rho }_{1}-{\bf \rho }_{2})\text{ },  \eqnum{16}
\end{eqnarray}
where ${\bf p}_{j}=-i\hbar \partial /\partial {\bf \rho }_{j},$ $%
V_{j}^{\prime }(\rho _{j})=V_{j}(z_{j}=0,\rho _{j})-E_{0,zj},$ $%
m_{j}^{\prime }(\rho _{j}>R)=m_{b,j},$%
\begin{eqnarray}
\frac{1}{m^{\prime }(\rho _{j}<R)} &=&\frac{1}{m_{w,j}}\int_{0}^{d/2}dz_{j}%
\left| \psi _{j}(z_{j})\right| ^{2}  \nonumber \\
&&+\frac{1}{m_{b,j}}\int_{d/2}^{\infty }dz_{j}\left| \psi _{j}(z_{j})\right|
^{2}\text{ },  \eqnum{17}
\end{eqnarray}
and the effective Coulomb interaction is 
\begin{equation}
V_{c}^{\prime }({\bf \rho })=-\frac{e^{2}}{\epsilon }\int_{-\infty }^{\infty
}dz_{1}dz_{2}\frac{\left| \psi _{1}(z_{1})\right| ^{2}\left| \psi
_{2}(z_{2})\right| ^{2}}{\sqrt{(z_{1}-z_{2})^{2}+\left| {\bf \rho }\right|
^{2}}}\text{ .}  \eqnum{18}  \label{eq23}
\end{equation}
Using a Gaussian shape for the longitudinal wave function of the
groundstate, the authors of Ref.~\cite{price} have obtained an analytical
approximation to the effective Coulomb potential 
\begin{equation}
V_{c}^{\prime }(\rho )=-\frac{e^{2}}{\epsilon }\frac{1}{\sqrt{2\pi }\gamma }%
e^{\rho ^{2}/4\gamma ^{2}}K_{0}(\frac{\rho ^{2}}{4\gamma ^{2}})\text{ ,} 
\eqnum{19}  \label{eq24}
\end{equation}
where $K_{0}$ is the modified Bessel function. For a system with infinite
barriers the value $\gamma =0.277d$ gives the best fit to the effective
Coulomb potential. As a consequence of the penetration of the electron and
hole into the barrier region, the value of $\gamma /d$ increases with
decreasing disk thickness. We have found that for our parameters of the
quantum disk, $R=8.95nm$ and $d=3.22nm,$ the value $\gamma =1.675nm$ gave
the best fit to the results obtained from direct numerical calculation of
the effective two-dimensional Coulomb potential. The other parameters of the
two-dimensional Hamiltonian are $E_{0,ze}=116.06meV,$ $E_{0,zh}=38.13meV,$ $%
V_{oe}^{\prime }=141.94meV,$ $V_{oh}^{\prime }=133.87meV,$ $m_{we}^{\prime
}=0.080m_{0},$ $m_{be}^{\prime }=0.097m_{0},$ $m_{wh}^{\prime
}=m_{bh}^{^{\prime }}=0.45m_{0},$ where indices $e,h$ correspond to electron
and hole, respectively. The numerical diagonalization technique for the 2D
Hamiltonian was presented already in Ref. \cite{peeters}.

\section{Results and discussion for the exciton groundstate}

We have calculated the exciton groundstate energy and exciton binding energy
as a function of an applied magnetic field. We used for our simulations the
physical parameters of the $In_{0.55}Al_{0.45}As$ self-assembled quantum
dots, used in the experiment by Wang {\it et al.} \cite{wang}. The studied
disks have a height of $3.22nm$ and a radius of $8.95nm.$ The other
parameters were already given above. Fig.~1 shows the probability
distribution of the electron (solid curves) and hole (dashed curves) $\left|
\psi _{i}({\bf r}_{i})\right| ^{2},$ $i=e,h,$ for the ground state along $%
\left( z=0,\rho \right) $ and perpendicular $\left( z,\rho =0\right) $ to
the disk under consideration. Along the $\rho $-direction, the electron and
hole are confined within the disk but along the $z$-direction, there is
appreciable penetration into the barrier material.

In Fig.~2 the exciton groundstate energy is plotted as a function of the
magnetic field. This groundstate energy is given by 
\begin{equation}
E_{0}=E^{e}+E^{h}+E_{exc}\text{ ,}  \eqnum{20}  \label{eq25}
\end{equation}
where $E^{e}$ and $E^{h}$ are the single electron and hole energies,
respectively and $E_{exc}$ is the exciton binding energy. The solid curve
shows the result of the full 3D treatment of the problem, whereas the dashed
and dotted curves are calculated using the adiabatic approximation. For the
latter case, we make a distinction between the cases with and without
correlation. For the case without correlation, the Coulomb interaction is
calculated using the single particle electron and hole wave functions 
\begin{equation}
E_{exc}=-\frac{e^{2}}{\epsilon }\left\langle \psi ^{e}\psi ^{h}\left| \frac{1%
}{\left| {\bf r}_{e}-{\bf r}_{h}\right| }\right| \psi ^{e}\psi
^{h}\right\rangle \text{ .}  \eqnum{21}  \label{eq26}
\end{equation}
The total exciton wave function was used in order to calculate the energy
with correlation 
\begin{equation}
E_{exc}=-\frac{e^{2}}{\epsilon }\left\langle \psi ^{e,h}\left| \frac{1}{%
\left| {\bf r}_{e}-{\bf r}_{h}\right| }\right| \psi ^{e,h}\right\rangle 
\text{ .}  \eqnum{22}  \label{eq27}
\end{equation}
Figure~2 shows an enhancement of the groundstate energy with increasing
magnetic field for all three cases. The correlation energy, which is given
by the difference between the dotted and dashed curve, is $3.4meV$ for $B=0T$
and increases to $4.4meV$ for $B=40T.$

The inset shows the exciton binding energy as a function of the magnetic
field. Again we see an increase for increasing magnetic field as expected.
This is not surprising, because by applying higher magnetic fields the
particles are more confined, they are closer to each other and therefore
more tightly bound, which implies an increase of the binding energy. The
assignment of the different curves is the same as for the groundstate energy
(see main figure). Note that the inclusion of correlation increases the
binding energy at $B=0T$ with $14.5\%$ while the full 3D treatment of the
problem further increases the binding energy with $13.6\%$.

From our calculation of the exciton groundstate energy, we can easily
determine the diamagnetic shift of the exciton, which is defined by $\Delta
E=E(B)-E(B=0).$ The result is shown in Fig.~3, where the curves indicate our
calculated results for the three cases, as mentioned above, and the squares
are the experimental results, as obtained by Wang {\it et al.} \cite{wang}.
From the comparison between the different approaches and experiment, we
notice: 1) for $B<8T$ all three approaches give practically the same result
which agrees perfectly with experiment, 2) when $B$ is increased above 8$T$
the three theoretical approaches have the same qualitative $B$-dependence
but there are small quantitative differences in the slope of the curves, and
3) in the high field regime, i.e. $B>20T,$ our theoretical results
substantially underestimate the experimental result. The masses used for
these calculations were the ones given by Wang {\it et al. }in Ref.~\cite
{wang} $\left( m_{w,e}=0.076m_{0},\text{ }m_{b,e}=0.097m_{0},\text{ }%
m_{h}=0.45m_{0}\right) $ and it is clear that here the heavy-hole mass was
used. However, in Ref.~\cite{itskevich} it was argued that for a magnetic
field normal to the sample plane, the light hole mass should be used.
Because the dot height is much smaller than the dot radius, heavy hole
character is expected in the growth direction for the ground hole state and
light hole character for in-plane motion. Therefore, for $B$ normal to the
sample plane, the light hole mass should be used. Including the effects of
strain, they find for $InAs$ dots that $m_{e}=0.055m_{0}$ and $%
m_{h}=0.1m_{0}.$ Combining this with values for $AlAs$ \cite{grundmannstier}%
, we find by linear interpolation to the material $In_{0.55}Al_{0.45}As$
values of $0.080m_{0}$ and $0.2m_{0}$ for respectively the electron and the
hole mass. The result for the diamagnetic shift in this case is depicted as
the dot-dashed curve in Fig.~3. This result is in very good agreement with
the experimental results.

In previous theoretical work, only the exciton energy and wave function at $%
B=0$ were considered, from which the diamagnetic shift can be calculated as $%
\Delta E=\beta B^{2},$ where $\beta =e^{2}\left\langle \rho
^{2}\right\rangle /8\mu $ and $\left\langle \rho ^{2}\right\rangle $ is the
mean quadratic electron-hole distance. This is a good approximation in case
of low magnetic fields, when the magnetic confinement is much lower than the
confinement due to the quantum dot. However, for higher magnetic fields, the
magnetic confinement becomes more important. Then the quadratic dependence
of the energy shift on the magnetic field will change into a linear
dependence, due to the formation of Landau levels. In this case the energy
shift becomes $\Delta E=\hbar (\omega _{c,e}+\omega _{c,h}),$ where $\omega
_{c,i}=eB/m_{i}$ is the cyclotron frequency. With this knowledge, one can
construct the function 
\begin{equation}
\Delta E=\frac{\beta B^{2}}{1+\alpha B}\text{ ,}  \eqnum{23}  \label{eq28}
\end{equation}
which interpolates between the small and large magnetic field behaviour and
where $\beta $ and $\alpha $ are taken as fitting parameters. This formula
gives for low magnetic fields $(B\rightarrow 0)$ the already known
expression $\Delta E=\beta B^{2},$ and for high magnetic fields $%
(B\rightarrow \infty )$ $\Delta E=(\beta /\alpha )B.$ It turns out that Eq.~(%
\ref{eq28}) gives an extremely good fit to the numerical results of Fig.~3
for $\beta =6.63\mu eVT^{-2}$ and $\alpha =3.25\times 10^{-3}T^{-1}.$ We
found that the fitted curve reproduces the solid curve in Fig.~3 so well
that they can not be discriminated. We also calculated $\beta $ using the
expression $\beta =e^{2}\left\langle \rho ^{2}\right\rangle /8\mu $ for $%
B\rightarrow 0,$ which resulted into the value $\beta =9.58\mu eVT^{-2}.$
This value is substantially higher than the one found by fitting. In the
other limit, we compare $\Delta E=(\beta /\alpha )B$ with $\Delta E=\hbar
\omega _{c}=(\hbar e/\mu )B,$ where $\mu =m_{e}m_{h}/(m_{e}+m_{h})$ is the
effective exciton mass in $InAlAs$. Such a calculation gives $\hbar e/\mu
=1.68\times 10^{-3}eVT^{-1},$ which is smaller than the fitted value $\beta
/\alpha =2.04\times 10^{-3}eVT^{-1}.$ The fitted results within the
adiabatic approximation with and without correlation are respectively, $%
\beta =7.56\mu eVT^{-2},$ $\beta /\alpha =1.52\times 10^{-3}eVT^{-1}$ and $%
\beta =7.79\mu eVT^{-2},$ $\beta /\alpha =2.75\times 10^{-3}eVT^{-1}$. Using
the light hole mass instead of the heavy hole mass, we find respectively the
following fitted and calculated results: $\beta =9.16\mu eVT^{-2},$ $\beta
/\alpha =2.27meVT^{-1}$ and $\beta =14.08\mu eVT^{-2},$ $\beta /\alpha
=1.96meVT^{-1}.$

In the above calculations we investigated the adiabatic shift, which is a
relative quantity, and therefore in the calculation of the groundstate
energy, the bandgap was not included. But when we want to compare the
experimental excitation energy, the bandgap of the disk material is needed.
For $B=0T$, using the heavy hole mass, we found a groundstate energy of $%
E=E^{e}+E^{h}+E_{exc}=152.2meV,$ using Eq.~(\ref{eq25}). For the case of the
light hole mass, which gave a better agreement with the experimental
results, the groundstate energy at $B=0T$ is $178.5meV.$ To obtain the total
excitation energy, as measured in photoluminescence experiments, e.g. by
Wang {\it et al.} \cite{wang}, the bandgap energy $E_{g}$ has to be added to
this equation: 
\begin{equation}
E=E^{e}+E^{h}+E_{exc}+E_{g}\text{.}  \eqnum{24}  \label{eq29}
\end{equation}
For our study, we considered $In_{0.55}Al_{0.45}As/Al_{0.35}Ga_{0.65}As$
quantum dots. Without strain, the bandgap energy of the dot material was
obtained by linear interpolation between the result for $InAs$ ($%
E_{g}=0.41eV $) and $AlAs$ ($E_{g}=3.13eV$) which results into $%
E_{g}=1.634eV,$ whereas we found for the barrier material that $%
E_{g}=2.083eV $ \cite{walle,krijn}$.$ The difference in bandgap between the
two materials is $\Delta E_{g}=450meV.$ For the total exciton energy, we now
find $E=1.634eV+0.152eV(0.1785eV)=1.786eV(1.8125eV)$ using respectively the
heavy (light) hole mass. From Ref. \cite{wang}, we know that the bandgap
difference between the dot and the barrier material, corrected for strain
effects, is $\Delta E_{g}=430meV.$ This means that the bandgap of the dot
material has increased with 20$meV.$ For the total exciton energy, this
gives us the final result of $E=1.81eV$ using the heavy hole mass and $%
E=1.83eV$ using the light hole mass.\ In the experiments, for $B=0T,$ the
value of $E=1.894eV$ was found, which gives a reasonable agreement with our
theoretical result in view of the fact that the composition of the alloy in
the dot can, for example, not be uniform, the dot size is not known with
high accuracy, etc.

Next, we investigated the effect of an applied magnetic field on the exciton
characteristics, using the parameters corresponding to the solid curve in
Fig.~3. First we considered the one-particle characteristics $\left\langle
z_{e}^{2}\right\rangle ^{1/2},$ $\left\langle z_{h}^{2}\right\rangle ^{1/2},$
$\left\langle \rho _{e}^{2}\right\rangle ^{1/2}$ and $\left\langle \rho
_{h}^{2}\right\rangle ^{1/2},$ where $z_{e}$, $z_{h}$ and $\rho _{e},$ $\rho
_{h}$ are the electron and hole coordinates along the $z$-axis and in the
plane, respectively. The results are shown in Fig.~4 and were calculated
using the full 3D approach. The figure shows clearly the squeezing of the
exciton due to the magnetic field, especially for the in-plane direction.
The mean quadratic electron-hole separations $\left\langle \rho
_{eh}^{2}\right\rangle ^{1/2}$ and $\left\langle z_{eh}^{2}\right\rangle
^{1/2}$ give an idea of the size of the exciton. We defined $\rho
_{eh}=\left| \overrightarrow{{\bf \rho }}_{e}-\overrightarrow{{\bf \rho }}%
_{h}\right| $ and $z_{eh}=\left| z_{e}-z_{h}\right| .$\ Notice that the size
of the exciton is comparable to the disk size. We see a more substantial
decrease with increasing magnetic field than for the single particle
wavefunction, which agrees with the increased binding of the exciton.

In Fig.~5 the percentage of the electron (right scale in Fig.~5) and hole
(left scale in Fig.~5) wavefunction in the dot is shown with varying
magnetic field. Both the results for the 2D case with correlation and the
full 3D treatment were calculated. More than $90\%$ of the hole is inside
the dot while only $71$-$73\%$ of the electron is inside the dot. With
increasing magnetic field both the electron and hole become more confined
inside the dot, indicating further the squeezing due to the magnetic field.
For the hole, we observe a flattening of the curve at very high magnetic
fields, both for the 3D calculation and for the 2D case. For high magnetic
fields, the hole wavefunction is in the $\rho $-direction totally confined
in the dot. However, there is still some extent of the wave function outside
the dot in the $z$-direction. But since the magnetic field has almost no
influence on the $z$-direction, applying higher magnetic fields will not
attribute to a further increase of the amount of the wavefunction inside the
dot and there will always be a small part of the wavefunction outside the
dot.

Figures~6(a,b) are contourplots of the density distribution, of the electron
and hole, respectively, along a cross section in the middle of the quantum
dot perpendicular to the $y$-direction. The electron\ density is defined as 
\begin{equation}
\left| \psi _{e}({\bf \rho }_{e},z_{e})\right| ^{2}=\int dz_{h}\int d{\bf %
\rho }_{h}\left| \psi ^{e,h}({\bf \rho }_{e},z_{e},{\bf \rho }%
_{h},z_{h})\right| ^{2}\text{ ,}  \eqnum{25}  \label{eq30}
\end{equation}
and similarly for the hole.\ The solid curves show the result for the case
of $B=0T,$ whereas the dashed curves are the result for $B=40T.$ The dashed
square indicates the position of the disk, which is only one fourth of the
actual disk size. Due to the magnetic field, we see an increase in the
density inside the dot, both for electron and hole. Along the $\rho $%
-direction the particles become more centered in the middle of the dot due
to the squeezing by the applied magnetic field. However in the $z$%
-direction, it seems at first sight that there is an expansion instead of
the expected squeezing, but a closer look (by normalizing the function to
its central value) tells us that this is not the case. Of course, the
magnetic field is applied along the $z$-axis and has no direct influence on
the exciton behaviour in the $z$-direction. In the $\rho $-direction
however, the magnetic field brings the electron and hole closer together.
This implies a stronger interaction and we expect that this effect should
also be seen in the $z$-direction. This is also the case for the mean
quadratic hole $\left\langle z_{h}^{2}\right\rangle ^{1/2}$ and
electron-hole separation $\left\langle z_{eh}^{2}\right\rangle ^{1/2}$ (see
Fig.~4), which decrease as a function of the magnetic field.

\section{Effect of changing the disk radius}

We investigated the effect of the size of the disk on the exciton binding
energy which is depicted in\ Fig.~7 for the case with the full 3D treatment,
for the parameters corresponding to the solid curve in Fig.~3. When varying
the disk radius from $R=1nm$ up to $R=15$ nm (the dot thickness was fixed to 
$d=3.22nm$), we see initially a strong increase of the exciton energy by
more than a factor 2 and beyond $R\simeq 2.5nm$ it decreases slowly for
increasing $R$. In the `large' $R$-region, the binding energy increases for
decreasing disk radius due to the larger confinement of the electron and
hole wavefunction. The electron and hole are forced to sit closer to each
other, which leads to an enhancement of the binding energy. This behaviour
continues until the disk radius reaches a value of $R\simeq 2.5nm,$ where
the binding energy reaches a maximum value of $E_{exc}=47meV$. The decrease
in the binding energy with decreasing $R$ is due to the fact that the
wavefunction of the particles start to spill over into the barrier material,
i.e. the electron and the hole become less confined, which leads to a much
smaller interaction and therefore a lower binding energy. This is due to the
competition between the confinement kinetic energy and the barrier material
potential energy. This is confirmed by Fig.~8, where the percentage of
electron and hole inside the dot is shown as function of the disk radius.
For $R>6nm$ these percentages increase very slowly with increasing $R.$ We
never reach $100\%$ because of the substantial penetration of the
wavefunction in the barrier material along the $z$-direction (the thickness
of the dot is only $d=3.22nm$). Note that for $R=1nm$ only $1.88\%$ of the
electron wavefunction is inside the dot but $24.20\%$ of the hole
wavefunction.

The effect of a magnetic field on the dot size dependence of the exciton
energy is also shown in Fig.~7 for the case of $B=40T.$ Notice that the
largest $B$-dependence is found for very small and very large $R.$ In both
situations the confinement of the electron and hole are smallest and
consequently the ratio between the magnetic energy and the confinement
energy is largest. For intermediate dot size, i.e. $3nm<R<7nm$ we observe
the smallest effect of a magnetic field on the exciton energy. This is the
region of dot size where the confinement potential is able to strongly
confine the electron and hole to a small region in space.

We also investigated the effect of varying $R$ on the electron-hole
separation, both in the $\rho $ and in the $z$-direction. Fig.~9 shows the
result for $\left\langle z_{eh}^{2}\right\rangle ^{1/2}$ and we see a rather
high starting value at $R=1nm,$ decreasing strongly for increasing $R.$ This
high value at small $R$ follows from the fact that a large part of the
wavefunctions is outside the dot, so the particles are not really confined
anymore, which means that they are farther away from each other. The inset
of Fig.~9 shows the electron-hole separation in the $\rho $-direction. Also
here, we start with a high value at very small $R,$ followed by a strong
decrease and a minimum of $\sqrt{\left\langle \rho _{eh}^{2}\right\rangle }%
=3.12nm$ at $R\simeq 3nm.$ Further increasing $R,$ we find again an
enhancement of $\left\langle \rho _{eh}^{2}\right\rangle ^{1/2},$ which
initially is linear in $R,$ but for $R>12nm$ starts to level off and reaches
a constant value in the limit $R\rightarrow \infty .$

The low magnetic field diamagnetic coefficient $\beta =e^{2}\left\langle
\rho _{eh}^{2}\right\rangle /(8\mu ),$ can be obtained from the results of
Figs.~8 and 9, where $\mu $ is the effective exciton mass. The result is
shown in Fig.~10 and we see a similar behaviour as for the radial
electron-hole separation (inset of Fig.~9). When calculating $\beta ,$ we
took into account the variation of the effective exciton mass $\mu $ with
varying disk radius. The effective mass is defined as 
\begin{equation}
\frac{1}{\mu }=\frac{1}{m_{e}}+\frac{1}{m_{h}},  \eqnum{26a}  \label{eq31a}
\end{equation}
with 
\begin{equation}
\frac{1}{m_{e}}=\frac{1-P_{w}}{m_{e,b}}+\frac{P_{w}}{m_{e,w}},  \eqnum{26b}
\label{eq31b}
\end{equation}
where $m_{e,b}=0.097m_{0}$ and $m_{e,w}=0.076m_{0}$ are the effective
electron masses in respectively the barrier and the well and $P_{w}$ is the
probability to find the electron in the well. In Fig.~8 we showed that there
is a considerable change of $P_{w}$ for varying $R,$ and this will have an
effect on $m_{e}$ and $\mu .$ For the hole we have the same mass in and
outside the well, and therefore $m_{h}=0.45m_{0}$ is independent of $R.$ In
Fig.~11, the evolution of $m_{e}$ and $\mu $ is depicted as a function of $%
R. $ We see that for very small disks, where most of the wavefunction is
outside the dot, the value of $m_{e}$ converges to $m_{e,b}$ as expected.
For larger disk radii, this value decreases and for $R\rightarrow \infty ,$
it reaches the limit $m_{e}=0.0809m_{0},$ which is larger than $m_{e,w}$ due
to the penetration of the electron along the $z$-direction in the barrier
because of the small thickness of the disk.

\section{Exciton energy spectrum}

The higher radial excited states $(N\neq 0),$ for angular momentum $L=0$ are
calculated within the adiabatic approximation. The result for a disk with
radius $R=8.95nm$ and thickness $d=3.22nm$ is shown in Fig.~12(a), which
clearly shows the appearance of anti-crossing of levels for higher $N$
states and the energy scale for such states is also substantially larger
than for the angular momentum states. This anti-crossing is due to the fact
that we consider a fixed angular momentum $L$ for all states, which is a
conserved quantity. The states with fixed $L$ are non degenerate. Again we
considered cases of different disk radii and we observe an enhancement of
the anti-crossing for smaller disks (Fig.~12(b)) and a diminishing of the
anti-crossing for larger disks (Figs.~12(b) and 12(c)).

To study the anti-crossing more closely, we considered the disk with $%
R=8.95nm$ and $d=3.22nm$ (parameters corresponding to the dashed/dotted
curve in Fig.~3 were used). For the 2D problem, the radial part of the
exciton wave function for a fixed $L$ can be written as (see also Ref.~\cite
{peeters}) 
\begin{eqnarray}
\psi ({\bf \rho }_{e},{\bf \rho }_{h})
&=&\sum_{k=1}^{k=k_{n}}\sum_{n=1}^{n=k_{n}}\stackrel{l=l_{m}}{%
\mathrel{\mathop{\sum\nolimits^{\prime }}\limits_{l=-l_{m}}}%
}C_{kn}^{l}R_{k,(L+l)/2}(\rho _{e})  \eqnum{27} \\
&&\times R_{n,(L-l)/2}(\rho _{h})e^{il/2(\phi _{1}-\phi _{2})+iL/2(\phi
_{1}+\phi _{2})}\text{ ,}  \nonumber
\end{eqnarray}
where $k$ and $n$ correspond to the energy levels of the one particle
problem of electron and hole, respectively and $l$ is the relative angular
momentum. The sum $%
\mathop{\textstyle\sum}%
^{\prime }$ indicates that only even values of the relative angular momentum 
$l$ are taken when $L$ is even, and odd values otherwise. By studying the
values of the coefficients $C_{kn}^{l}$, we could distinguish which
one-particle states contribute most to the total exciton state. In
Figure~13(a), the symbols on the curves indicate which is the dominant term,
contributing to Eq.~(27). The inset in Fig.~13(b) gives the $(l,n)$ value
corresponding to the different symbols. Notice that $k$ remains 1, while $n$
can have higher values, which implies that the hole excited states are mixed
in into the exciton wave function. The single particle hole states have
lower energy, due to its higher effective mass. Notice that when we connect
each symbol by a line, we obtain intersecting levels. Such a spectrum would
be obtained if $(l,k,n)$ would be conserved quantities. Because of the
electron-hole interaction the different $(l,k,n)$ single particle states are
mixed which leads to the anti-crossing of the levels.\ Because $L$ is a
conserved quantity, crossings between different $N$ states are prohibited
and therefore, for a fixed $N,$ the system is forced to go to a different $%
\left( l,k,n\right) $-state.

The symbols in Fig.~13(a) indicate only the dominant term in Eq.~(27), while
the total summation considers typically about 750 terms. In Fig.~13(b) we
show how large the contribution of the dominant term is relative to the
total sum of all terms. This percentage is defined as 
\begin{equation}
\text{percentage }=\frac{\left| C_{kn}^{l}\right| ^{2}}{\sum\limits_{k,n,l}%
\left| C_{kn}^{l}\right| ^{2}}\times 100\text{.}  \eqnum{28}  \label{eq34}
\end{equation}
In Fig.~13(b) we only show the result for the case of $N=1,2,5$ in order not
to overload the figure. We want to emphasize that, as a function of the
magnetic field, the contribution of the dominant term, which can differ with
increasing magnetic field, is shown and not the evolution of the
contribution of a particular state. For $N=1,$ the $\left( 0,1,1\right) $%
-state appears to be very stable, as it stays between 85\% and 95\%. This
means that there is very little mixing with other states. The $\left(
0,1,2\right) $-state for $N=2$ is also very stable at low fields, but from $%
B=20T,$ the percentage drops, which indicates that another state is becoming
important and serious mixing occurs. Finally at $B=35T$ the $\left(
-1,1,1\right) $-state becomes most important, which can also be seen in
Fig.~13(a). Now the percentage of the contribution of this $\left(
-1,1,1\right) $-state is plotted and we see an increase with $B$. Also for $%
N=5,$ the transitions between the successive states are clearly visible from
Fig.~13(b). Each dip corresponds to a transition to another state with a
consecutive anti-crossing of the energy levels. A dip indicates strong
mixing between 2 or even 3 states where the dominant state gives only a
slightly higher contribution than the other important state(s). We see
strong dips at $B=12T$ and $B=20T,$ which indicate the mixing between
respectively $\left( 0,1,3\right) $ $\longleftrightarrow $ $\left(
1,1,1\right) $ and $\left( 1,1,1\right) $ $\longleftrightarrow $ $\left(
0,1,3\right) .$ Comparison with Fig.~13(a) shows that these magnetic field
values mark also the anti-crossings between respectively $%
N=4\longleftrightarrow N=5$ and $N=5\longleftrightarrow N=6.$ The height of
the peaks is an indication of the stability of the state. At $B=15T$ e.g.,
we see a very strong peak, whereas at $B=24T,$ only a small peak appears.
Comparing with Fig.~13(a) learns that, at the region between 20 and 30$T,$
there is a strong anti-crossing and, although, the system passes through the 
$\left( 0,1,3\right) $ state, this state never becomes really important.

In the above discussion, we indicated the states with $\left( l,k,n\right) ,$
where $l$ is an integer number. However, the relative angular momentum $l$
is not a good quantum number, and therefore the expectation value of the
operator $l_{z}=\left( \hbar /i\right) \partial /\partial \left( \phi
_{1}-\phi _{2}\right) $ is expected not to be an integer. In Fig.~14 the
expectation value $\left\langle l_{z}\right\rangle $ is depicted for the
different $N$-states as a function of the magnetic field, where $%
\left\langle l_{z}\right\rangle $ is calculated by 
\begin{equation}
\left\langle l_{z}\right\rangle =\sum_{l,k,n}l\left| C_{kn}^{l}\right| ^{2}%
\text{.}  \eqnum{29}
\end{equation}
Notice that $\left\langle l_{z}\right\rangle $ tends to approach an integer
value $l$ when one of the terms in the sum of Eq.~(27) dominates. The
transition between states with different $l$ is continuous. The more stable
a state is, the better it approaches an integer value $l$. The result for $%
N=7$ e.g. starts from $\left\langle l_{z}\right\rangle =0$ at $B=0T,$ then
increases up to about 0.75$\hbar $ and at $B=10T,$ drops down to $-1.8\hbar $
until $B=20T$ where it starts to increase again, more slowly now, up to $%
\left\langle l_{z}\right\rangle =0.9\hbar $ for $B=25T$ until finally at $%
B=40T$ it drops to less than $-2.5\hbar .$ This agrees very well with the
predicted integer values for $l$ in Fig.~13(a). For other $N$-states, the
agreement might be less good, which is due to the higher mixing with other
states.

Finally, we considered the energy states for different values of the total
angular momentum $L$ within the adiabatic approximation. The result for a
disk with radius $R=8.95nm$ and thickness $d=3.22nm$ is shown in Fig.~15(a)
for the lowest radial state $N=1$. Notice that for $B=0T,$ the states with $%
L $ and $-L$ are degenerate, which is lifted by a magnetic field. The
corresponding splitting is the well-known Zeeman splitting. For a smaller
disk radius, $R=5nm,$ all energies are shifted to higher values, the
splitting between the energy levels is larger, and the Zeeman splitting is
increased (Fig.~15(b)). When increasing the disk radius $R,$ i.e. $R=15nm$
and $R=30nm,$ the difference between the different angular momentum levels
decreases and the energy shifts to lower values (Figs.~15(c,d)). As in
previous case of different $N$-states, also here we studied which single
particle $\left( l,k,n\right) $-states are most important in the sum of
Eq.~(27) and the percentage of their contribution. Fig.~16(a) denotes, for a
disk with radius $R=8.95nm$ and thickness $d=3.22nm,$ the energies of the
different $L$-states and the symbols indicate which $\left( l,n\right) $%
-state is most important at a particular value of the magnetic field. Note
that here both $k$ and $n$ remain 1. In Fig.~16(b) the percentage
contribution of the particular $(l,k,n)$ state is depicted. We see a
transition occurring for the $L=-1,-2$ and $-3$ states. This follows also
from Fig.~17, where the expectation value $\left\langle l_{z}\right\rangle $
of the relative angular momentum operator is plotted. For the $L=0,1,2,3$
states, $\left\langle l_{z}\right\rangle $ remains quite constant, whereas
for the other $L$-states, $\left\langle l_{z}\right\rangle $ decreases
towards a lower value of $l.$ Because the total angular momentum is a
conserved quantity, energy levels corresponding to different $L$-values are
allowed to cross, they do not mix.

\section{Conclusions}

We calculated the groundstate energy (and the excited states), the binding
energy and the diamagnetic shift of an exciton in a quantum disk with radius 
$R$ and thickness $d$ for a hard wall confinement potential of finite
height. The mass mismatch between the dot material and the surrounding
material was taken into account. Our calculation is based on the finite
difference technique, where we used three different theoretical approaches,
which include the electron-hole correlation on different levels. The 3D
treatment is valid for arbitrary values of $R$ and $d$ and provides an
`exact numerical' treatment of the exciton problem. For $R\gg d,$ the
adiabatic approach is applicable and here we distinguish the cases with and
without correlation. The latter only uses the single particle wave functions
in order to calculate the exciton binding energy, whereas the first uses the
total exciton wavefunction.

Under the influence of an external applied magnetic field up to 40$T,$ we
find an increase of the exciton groundstate energy and binding energy. The
electron-hole separation shows a squeezing of the exciton due to the
magnetic field. This can also be seen from the electron and hole densities
in and around the dot. Our theoretical results of the diamagnetic shift are
in very good agreement with the experimental results of Ref. \cite{wang} if
we assume that the light hole is involved in the exciton.

When considering a varying disk radius $R,$ we found a strongly decreasing
exciton binding energy with decreasing $R$ for very small $R$-values, which
indicates that the dots are too small to confine the exciton. This
explanation is corroborated by an investigation of the radial electron-hole
separation and of the percentage of the wavefunction in the dot, which
indeed shows that, for very small $R,$ a large part of the wavefunction is
situated outside the dot. In the large $R$-regime the exciton binding energy
decreases with increasing $R$ and approaches a constant value for $%
R\rightarrow \infty .$ In the presence of an applied magnetic field, the
exciton binding energy approaches a constant value for large disks much
earlier than for the $B=0T$ case, indicating that the dot confinement is
dominated by the magnetic confinement.

Results for higher excited radial states, $N>0,$ show an anti-crossing of
levels which is more pronounced for small dot radius. The total angular
momentum $L$ is a conserved quantity. The relative angular momentum $l,$
however, is not a good quantum number. Because of the coupling between the
electron and the hole, the exciton wave function is a linear combination of
all possible one-particle wave functions. We investigated which $\left(
l,k,n\right) $-states contribute most and how large its contribution is to
the total exciton wavefunction. Furthermore we investigated the expectation
value of the relative angular momentum operator $l_{z},$ which is not
quantized and varies with the magnetic field. The degeneracy of the
different total angular momentum states is lifted due to the presence of the
confinement potential and the Zeeman splitting. This splitting decreases
with increasing dot radius $R.$ Also here we investigated the contribution
of the one particle states to the total exciton wavefunction. The energy
states with different total angular momentum $L$ can cross with varying
magnetic field, because $L$ is a good quantum number.

\section{Acknowledgments}

Part of this work is supported by the Flemish Science Foundation (FWO-Vl),
BOF-GOA and IUAP-IV. K. L. J. is supported by IWT, F. M. P. is a research
director with the FWO-Vl and V. A. S. was supported by a DWTC-fellowship.
Discussions with Dr. M. Hayne are gratefully acknowledged.

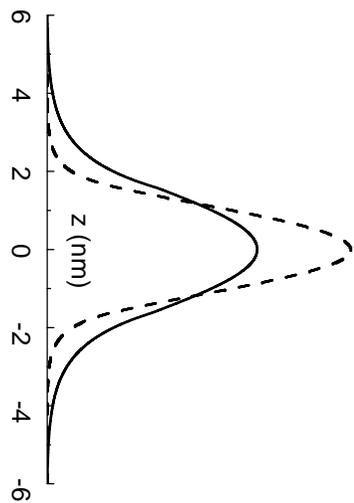
\begin{figure}[tbp]
\caption{Side view of the quantum disk together with the electron (solid
curves) and hole (dashed curves) probability distribution along the $\left( 
\protect\rho ,z=0\right) $ and the $\left( \protect\rho =0,z\right) $
direction.}
\label{Fig1}
\end{figure}

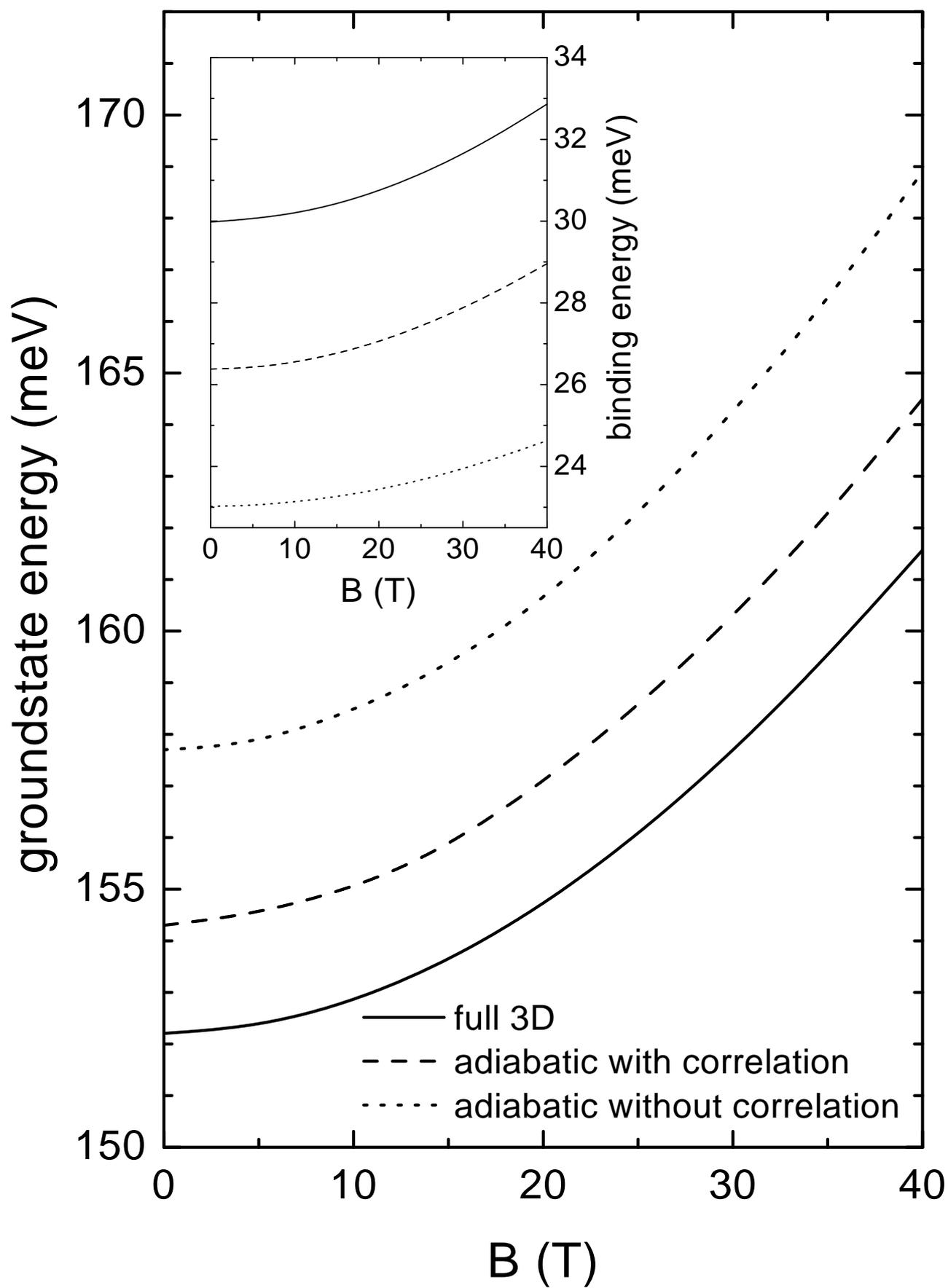
\begin{figure}[tbp]
\caption{The exciton groundstate energy as a function of the magnetic field.
The solid curve shows the result obtained within the full 3D treatment,
whereas the dashed and dotted curves are the result for the adiabatic
approximation, respectively with and without correlation. In the inset, the
exciton binding energy is plotted. The same curve conventions are used as in
the main figure. }
\label{Fig2}
\end{figure}

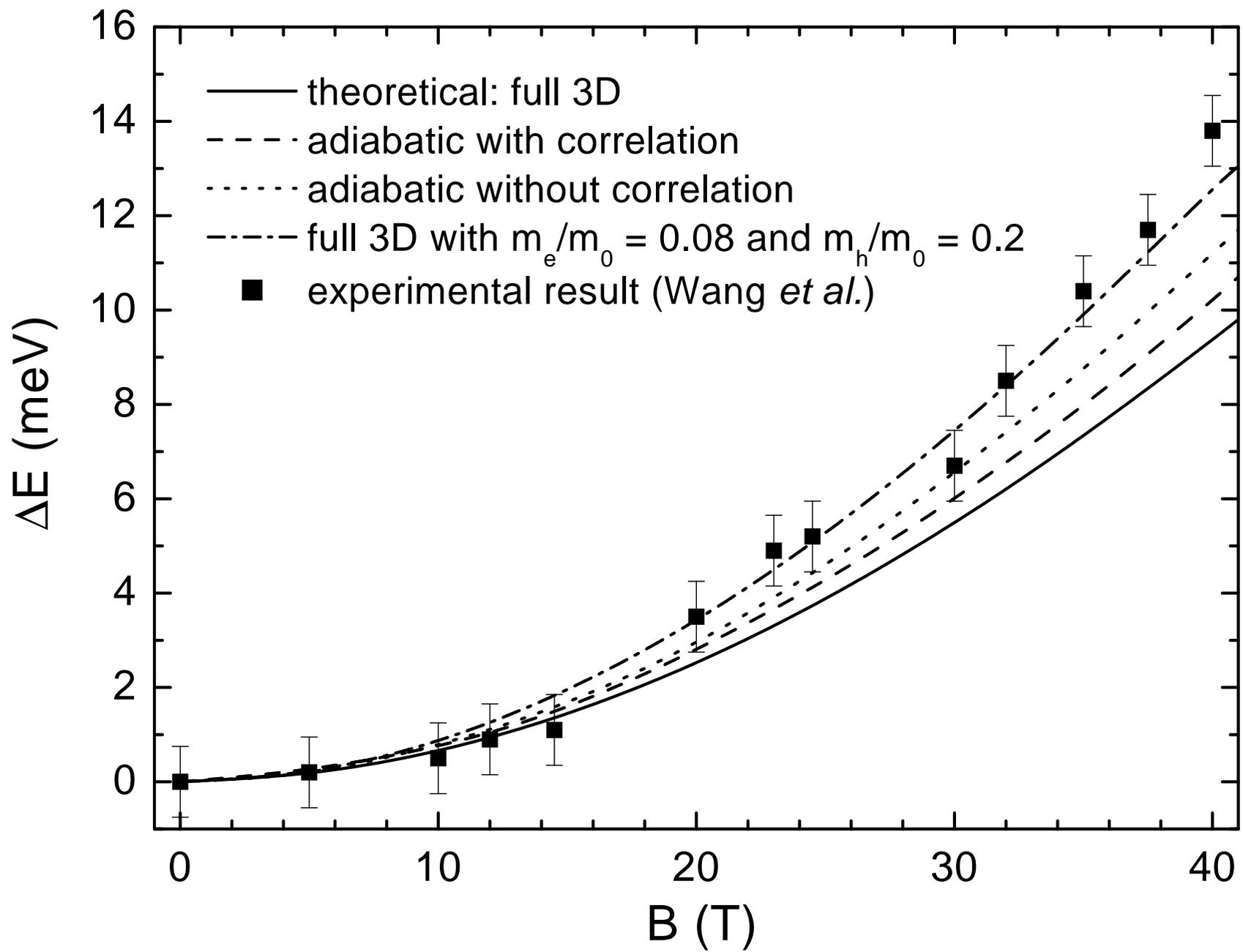
\begin{figure}[tbp]
\caption{The diamagnetic shift of the exciton energy as a function of an
external magnetic field. The curves are our theoretical results within
different approximations and the squares are the experimental results of
Wang {\it et al.} \protect\cite{wang}.}
\label{fig3}
\end{figure}

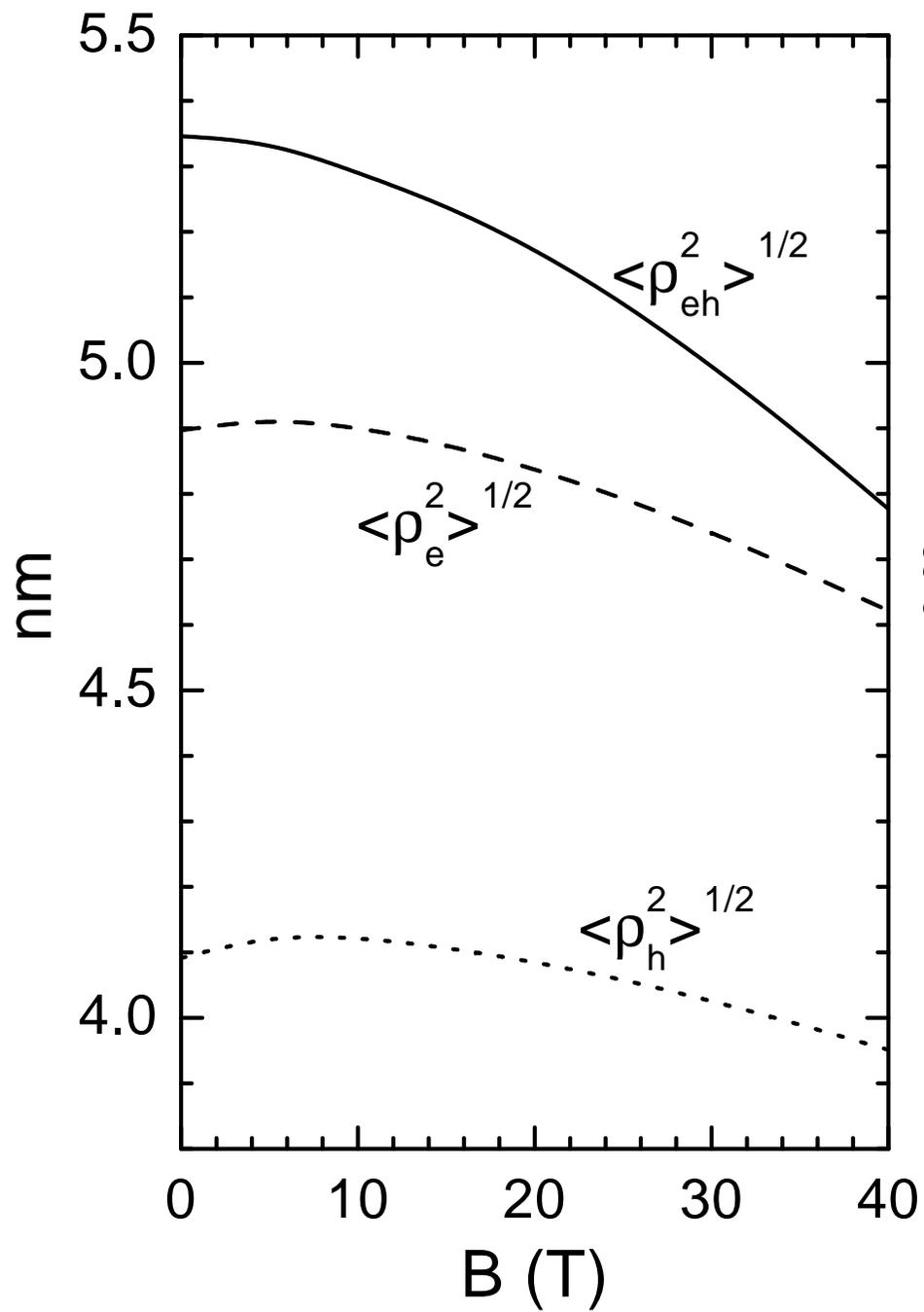
\begin{figure}[tbp]
\caption{The extent of the electron, the hole and the exciton as a function
of the magnetic field, along the radial (a) and longitudinal (b) direction,
respectively. }
\end{figure}

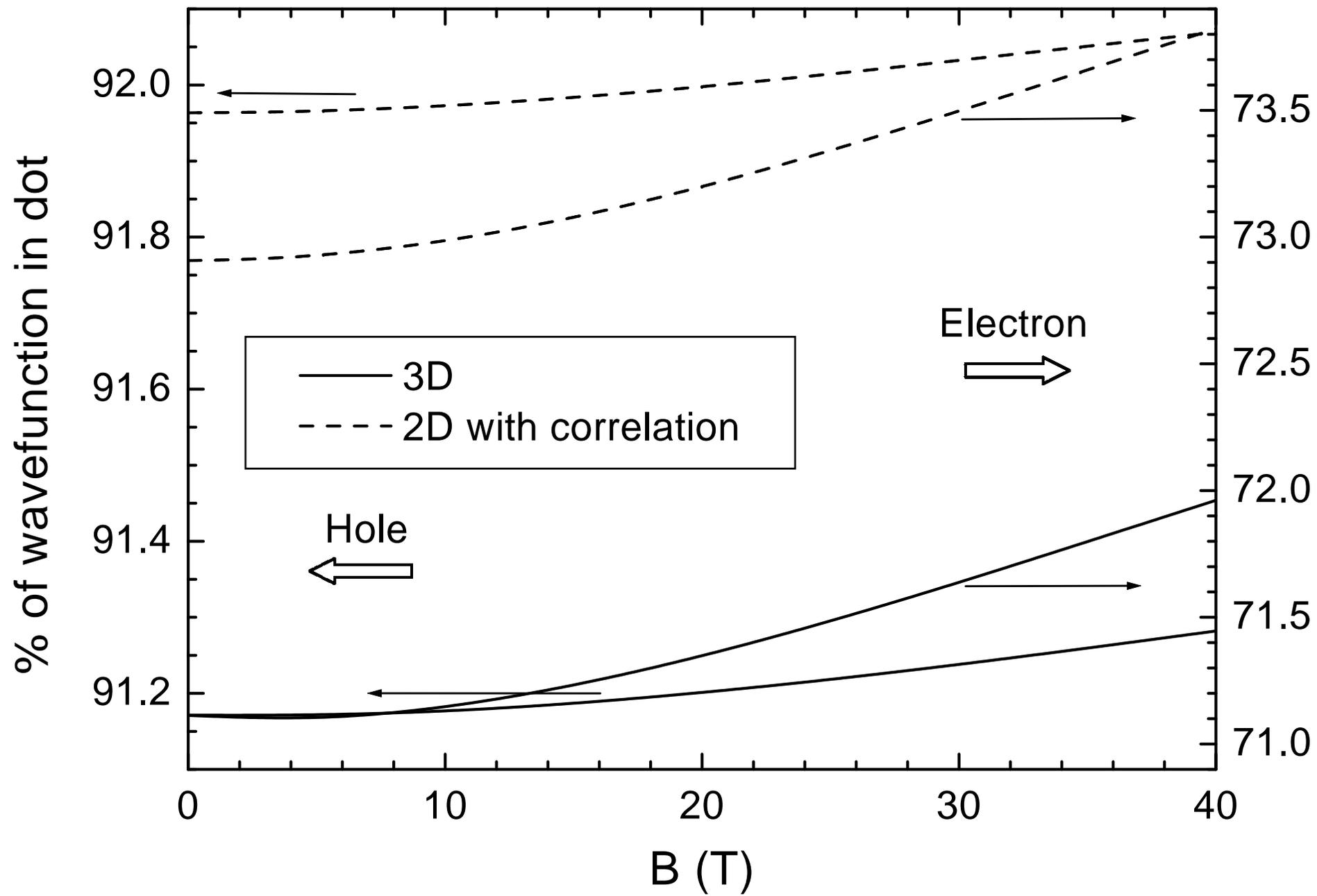
\begin{figure}[tbp]
\caption{The percentage of the wavefunction in the dot as a function of the
magnetic field, both for the electron (right axis) and the hole (left axis).
The solid curves are the result for the 2D case with correlation, the dashed
curves are the result obtained within the full 3D treatment.}
\label{fig5}
\end{figure}

\begin{figure}[tbp]
\caption{Contourplot of the (a) electron and (b) hole density in a plane
through the center of the quantum dot with size $R=8.95nm$ and $d=3.22nm.$
The plot shows only one quarter of the total space. Results are shown for $%
B=0T$ (solid curves) and for $B=40T$ (dashed curves).}
\label{fig6}
\end{figure}

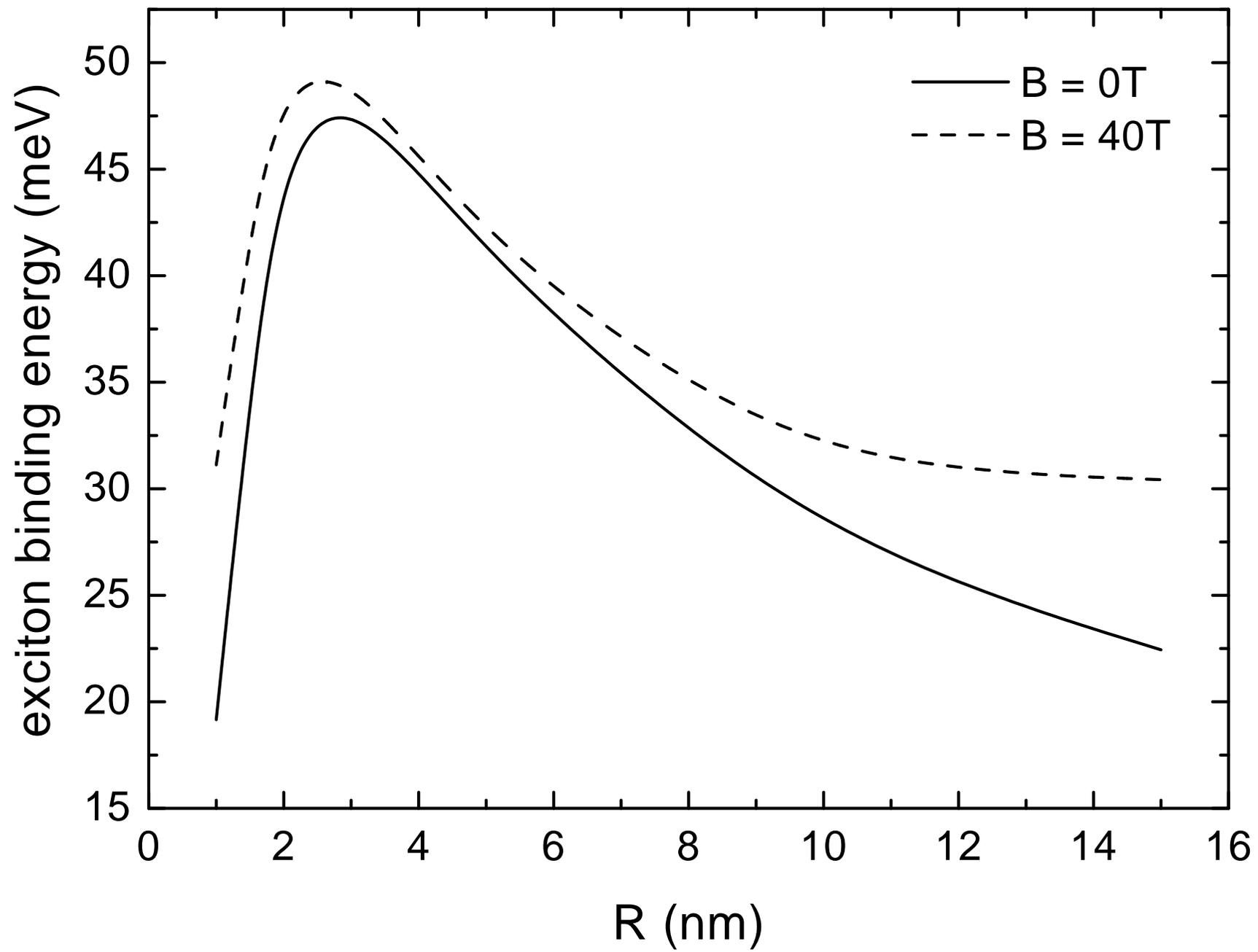
\begin{figure}[tbp]
\caption{The exciton binding energy as a function of the disk radius $R$ for
a disk thickness of $d=3.22nm$. Results are shown for two magnetic fields as
indicated.}
\label{fig7}
\end{figure}

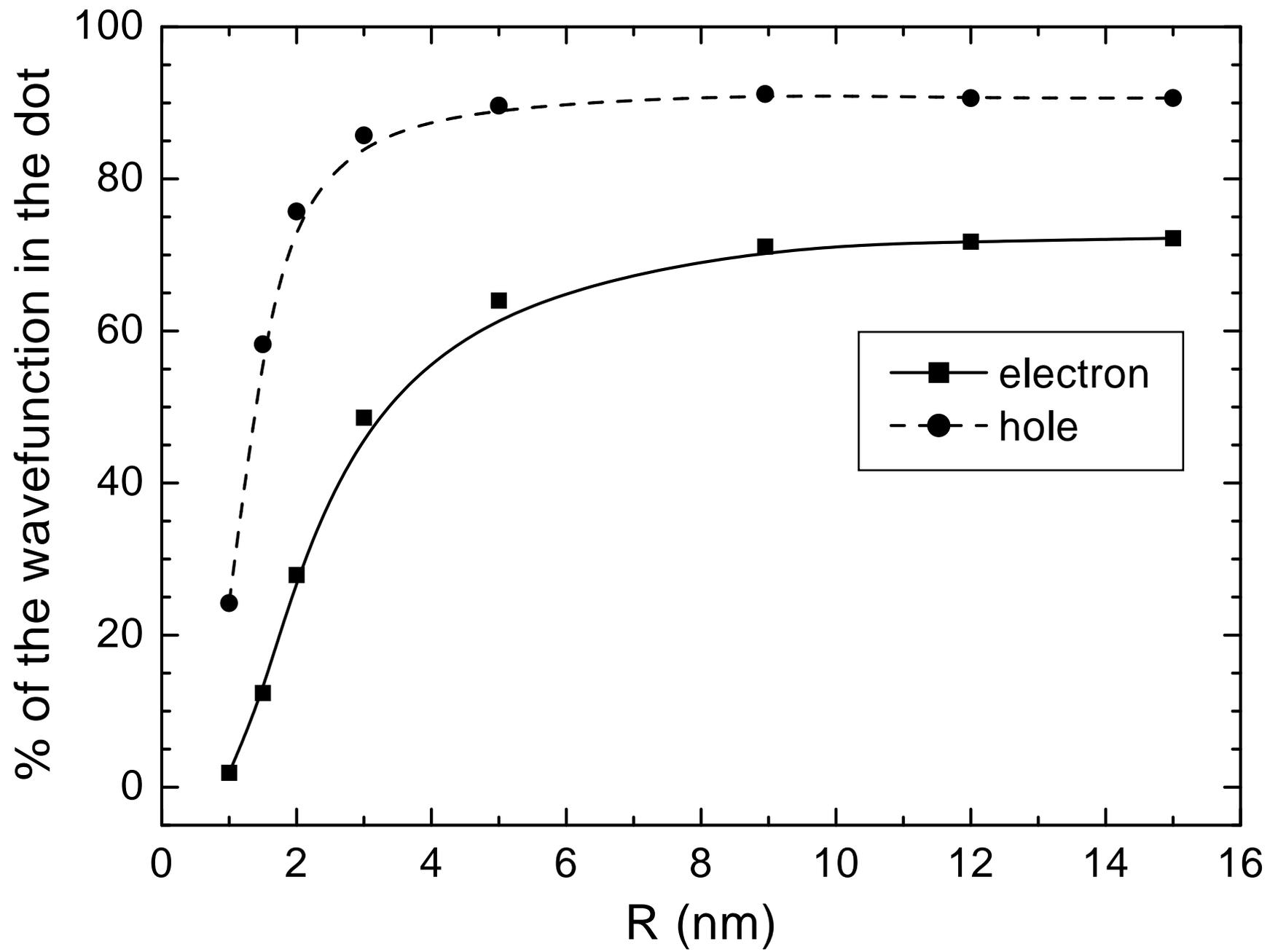
\begin{figure}[tbp]
\caption{The percentage of the electron (solid curve) and hole (dashed
curve) wavefunction in the dot as a function of the disk radius $R.$ Symbols
are the calculated points and the curve is a guide to the eye.}
\label{fig8}
\end{figure}

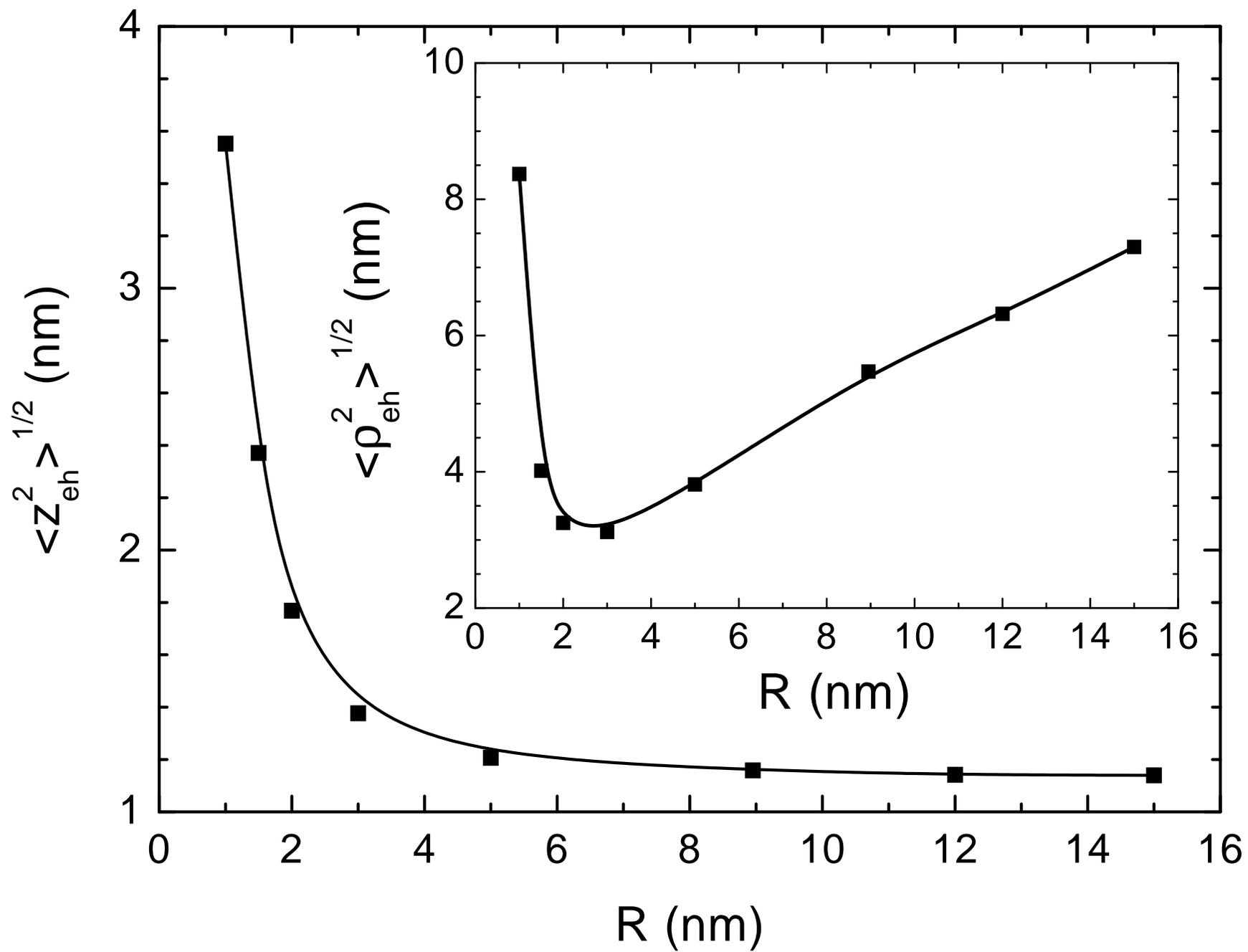
\begin{figure}[tbp]
\caption{Electron-hole separation in the $z$-direction as a function of $R.$
The inset shows the radial electron-hole separation as a function of $R.$}
\label{fig9}
\end{figure}

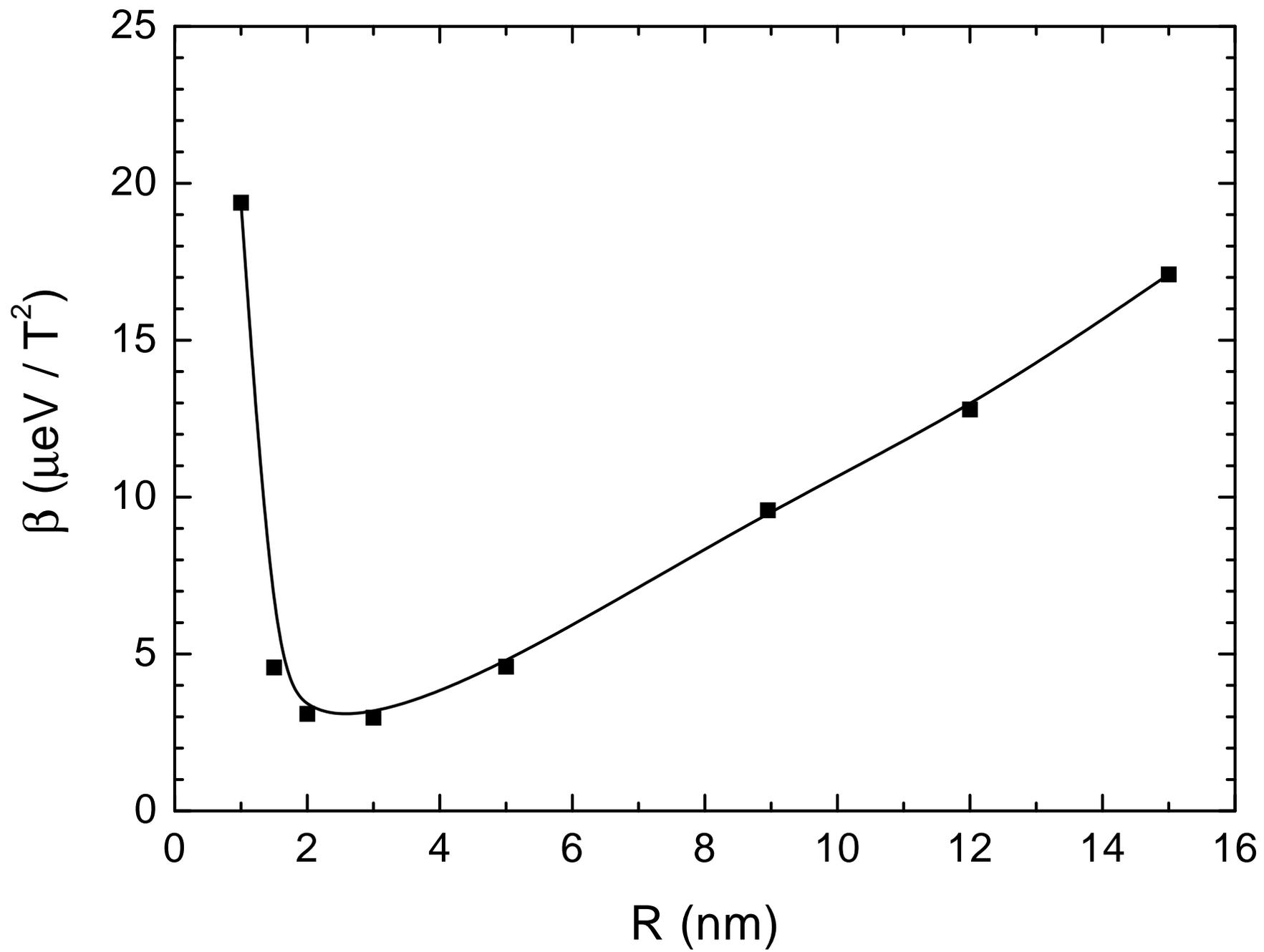
\begin{figure}[tbp]
\caption{The diamagnetic coefficient $\protect\beta $ as a function of the
disk radius $R$ for $d=3.22nm.$}
\label{fig10}
\end{figure}

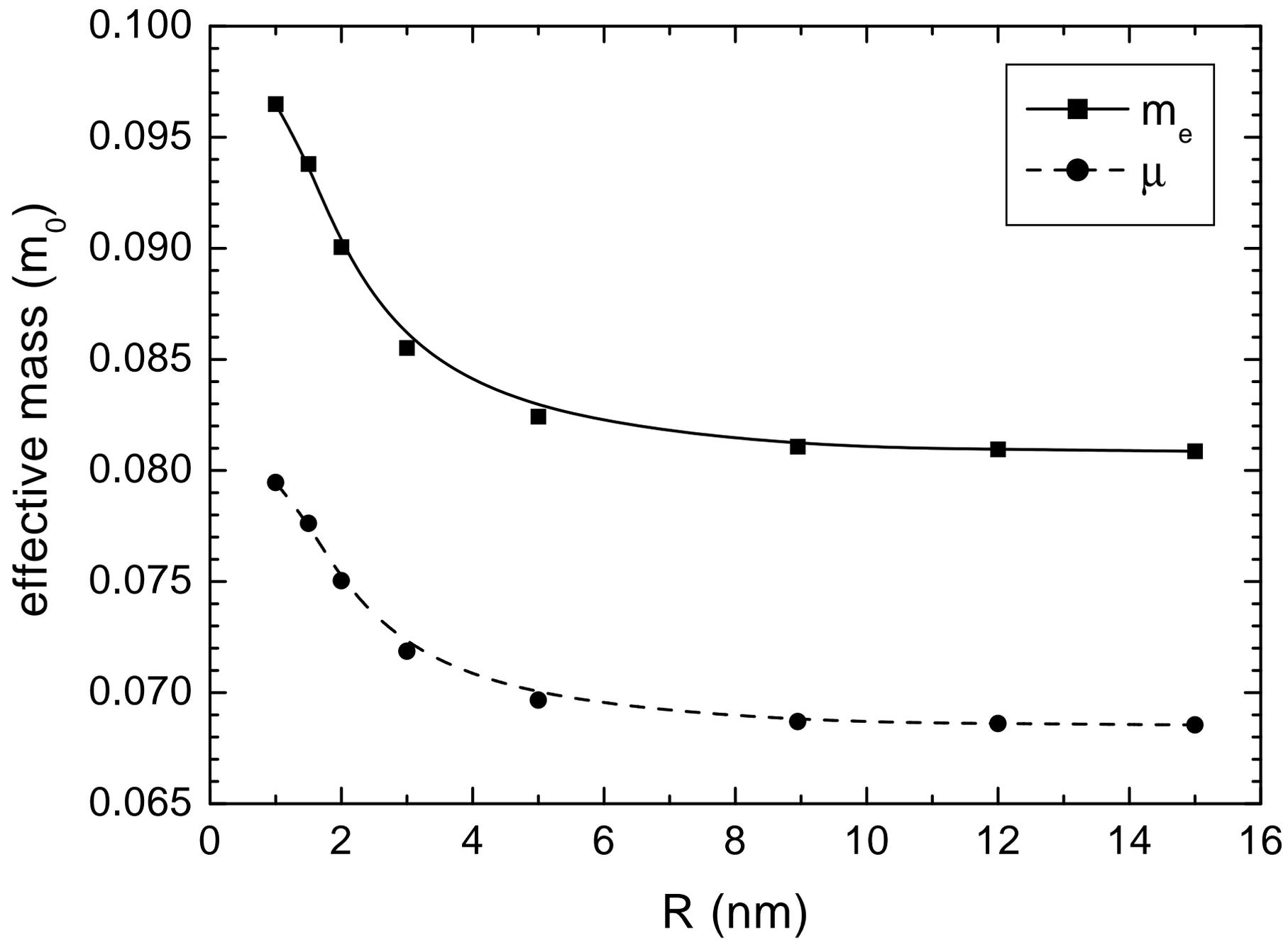
\begin{figure}[tbp]
\caption{The effective electron (solid curve) and exciton (dashed curve)
masses as a function of the disk radius $R.$}
\label{fig11}
\end{figure}

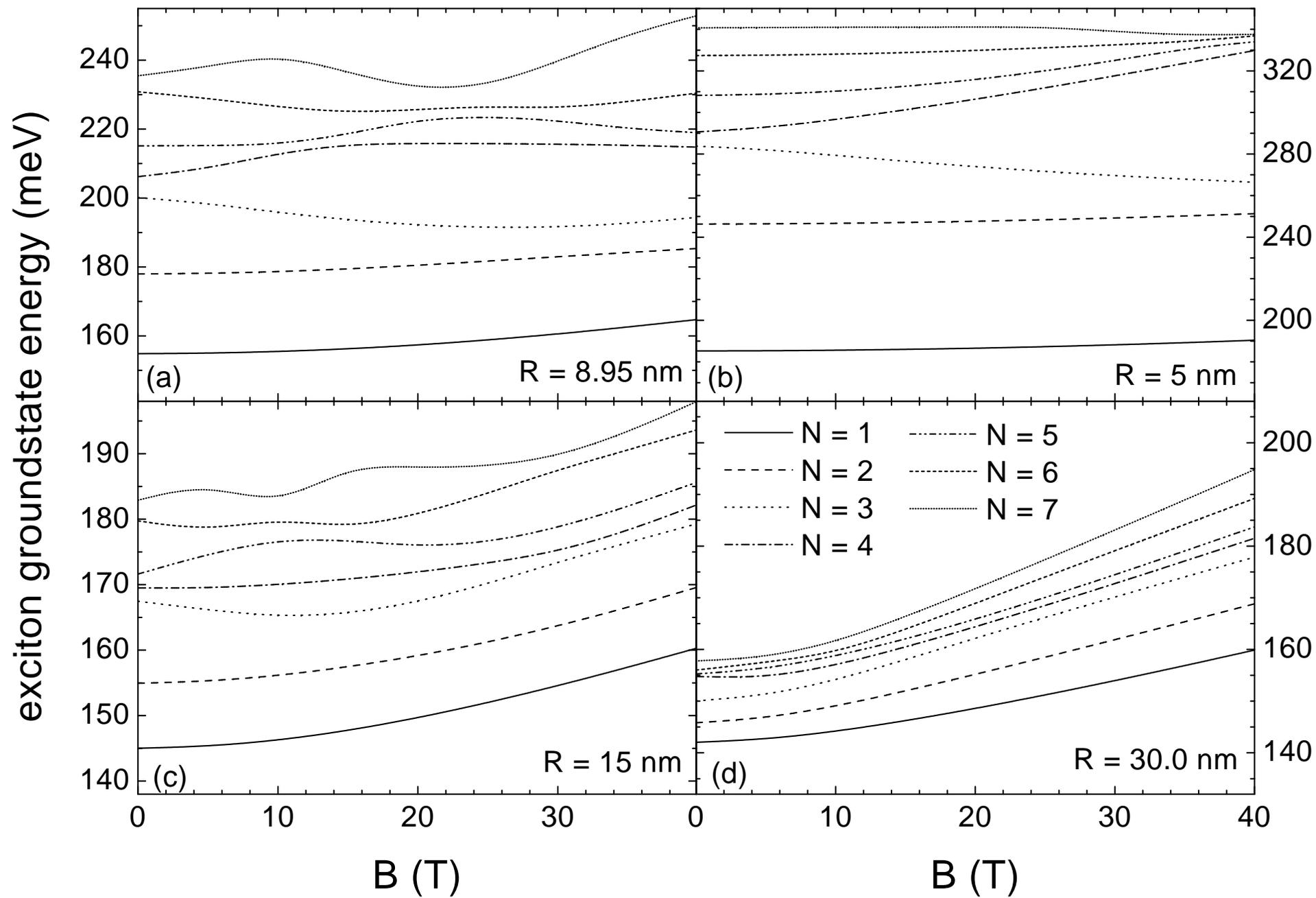
\begin{figure}[tbp]
\caption{The exciton energy states for different values of the radial
quantum number $N,$ as a function of the magnetic field for $L=0$. Four
different disk radii are considered: $R=8.95nm$ (a), $R=5nm$ (b), $R=15nm$
(c) and $R=30nm$ (d).}
\label{fig12}
\end{figure}

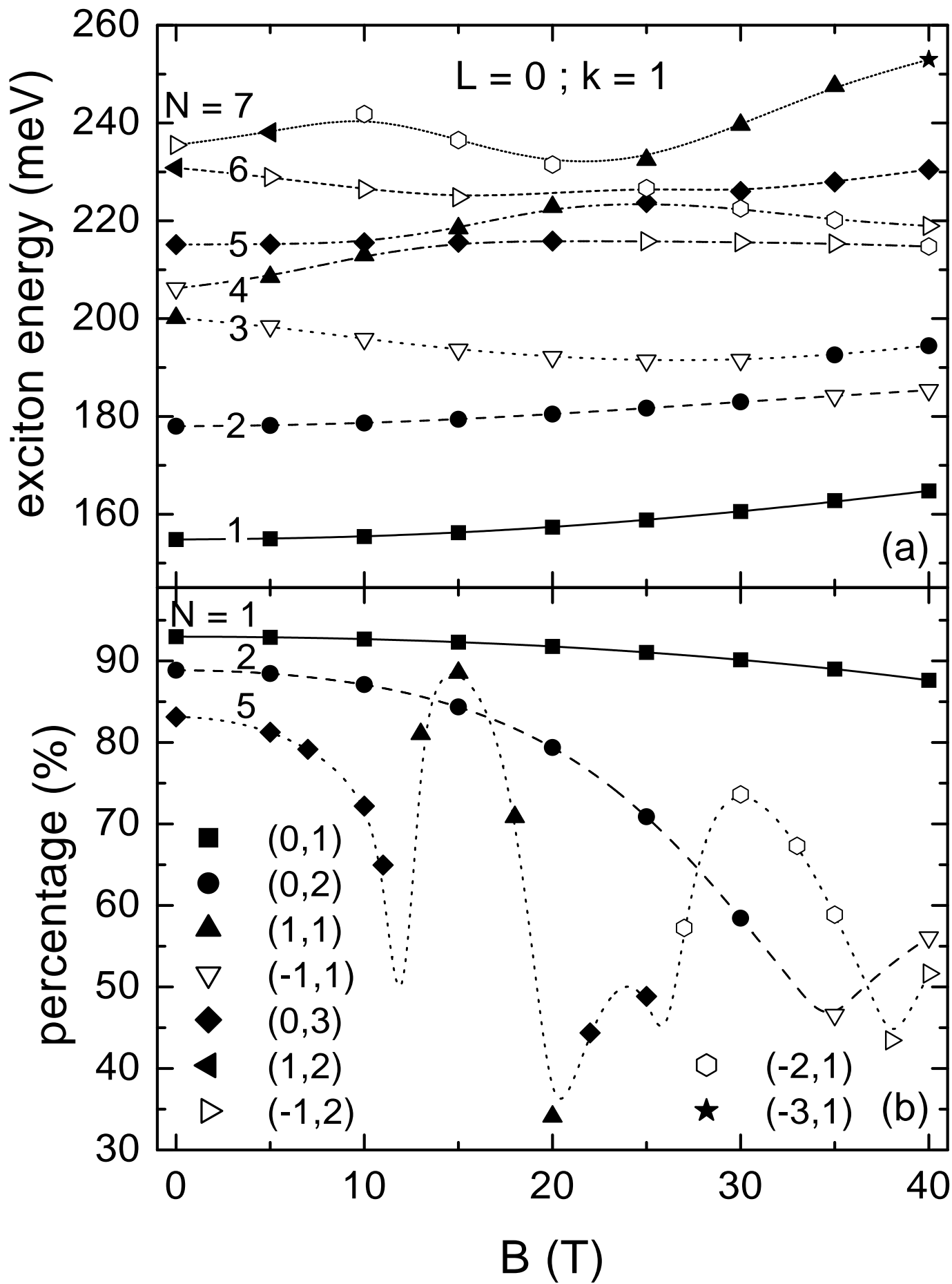
\begin{figure}[tbp]
\caption{(a) The higher excited radial states for $R=8.95nm$ and $d=3.22nm.$
The symbols indicate the $(l,n)$ value of the most important single-particle
state, where $l$ is the relative angular momentum and $n$ is the radial hole
quantum number. (b) The percentage of the contribution of the most important
state as a function of the magnetic field.}
\label{fig13}
\end{figure}

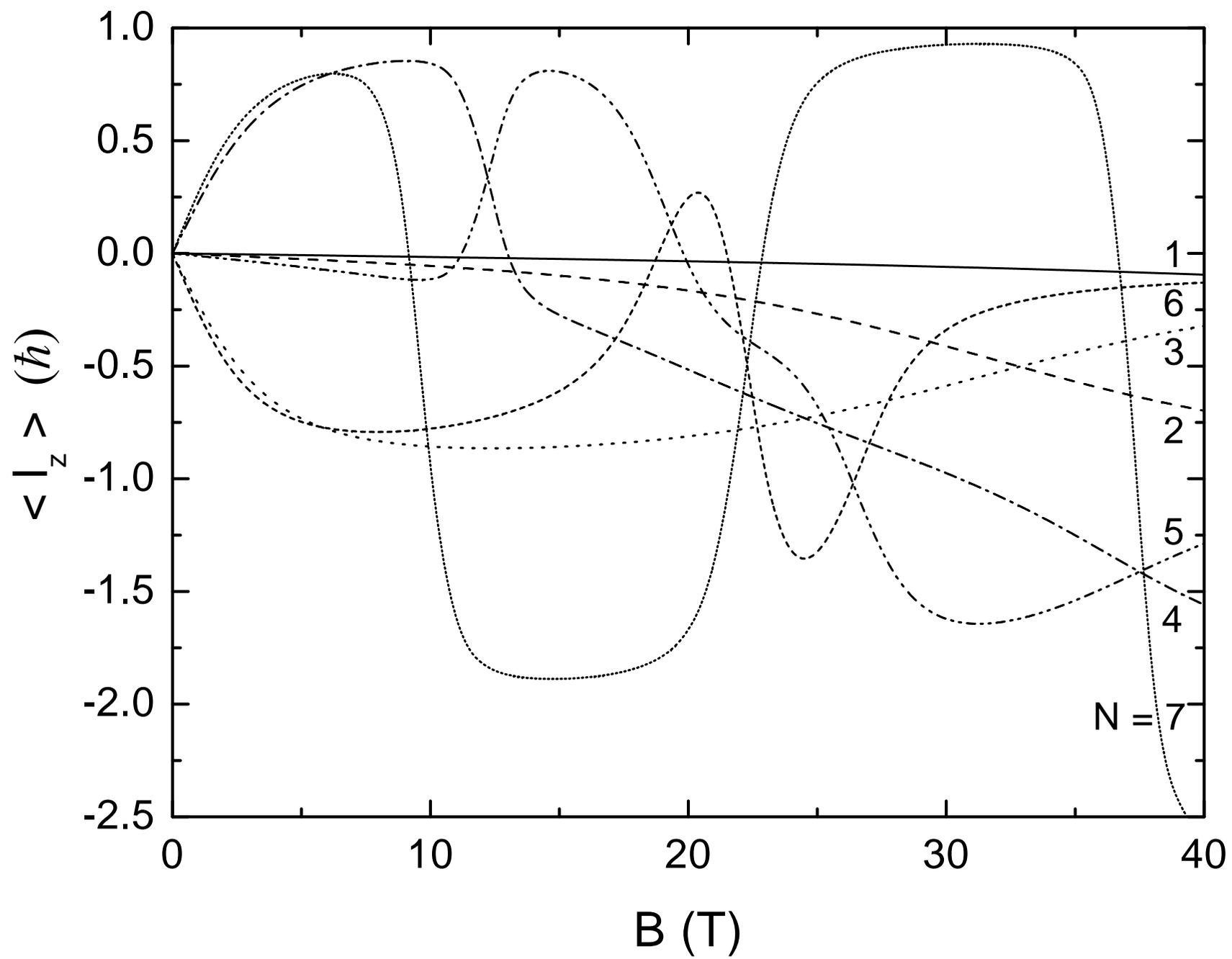
\begin{figure}[tbp]
\caption{The expectation value of the relative angular momentum operator $%
l_{z}$ as a function of the magnetic field, for $L=0$ and different $N$%
-states.}
\label{fig14}
\end{figure}

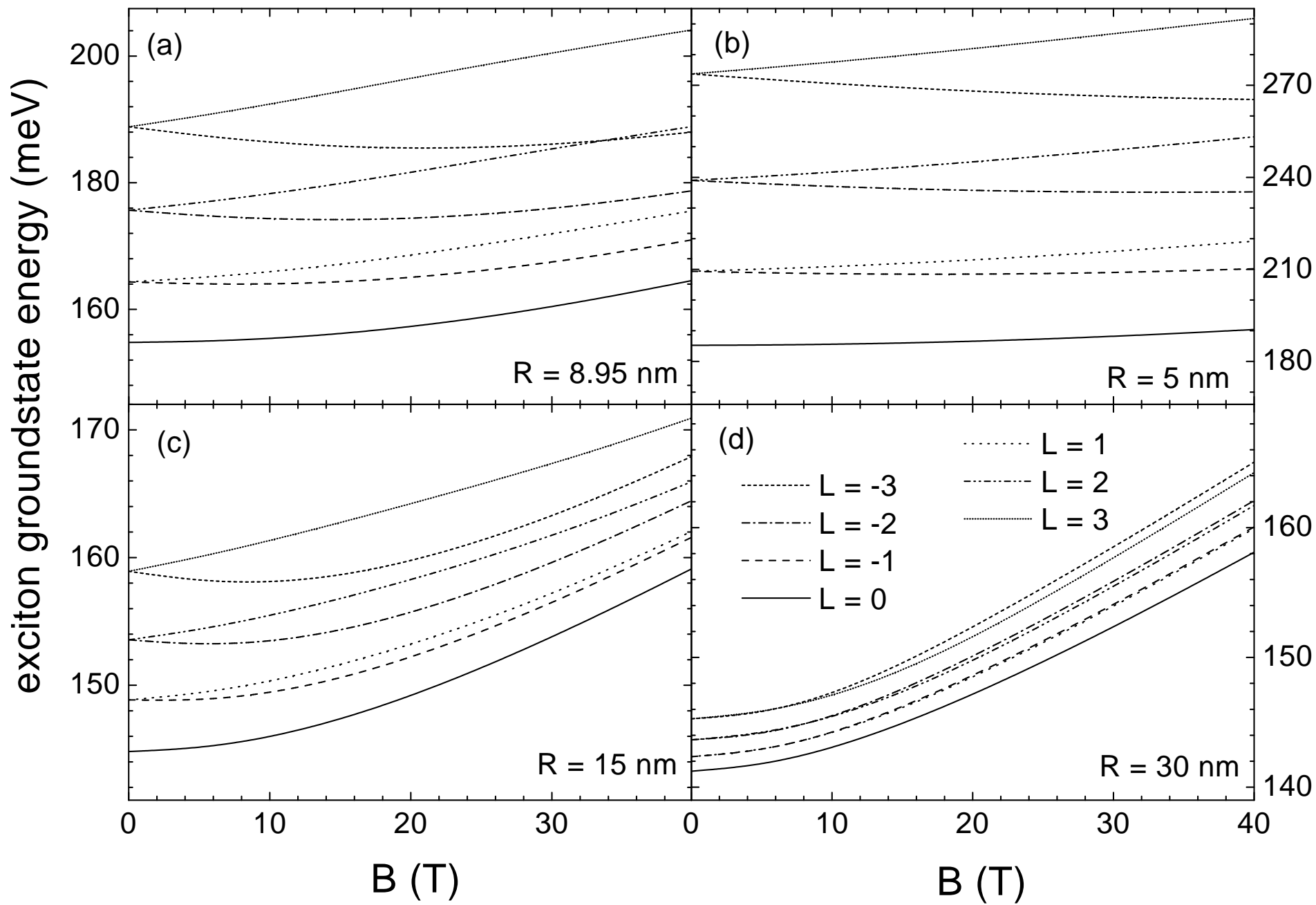
\begin{figure}[tbp]
\caption{The exciton energy states for different values of the total angular
momentum $L,$ as a function of the magnetic field. Four different disk radii
are considered: $R=8.95nm$ (a), $R=5nm$ (b), $R=15nm$ (c) and $R=30nm$ (d).}
\label{fig15}
\end{figure}

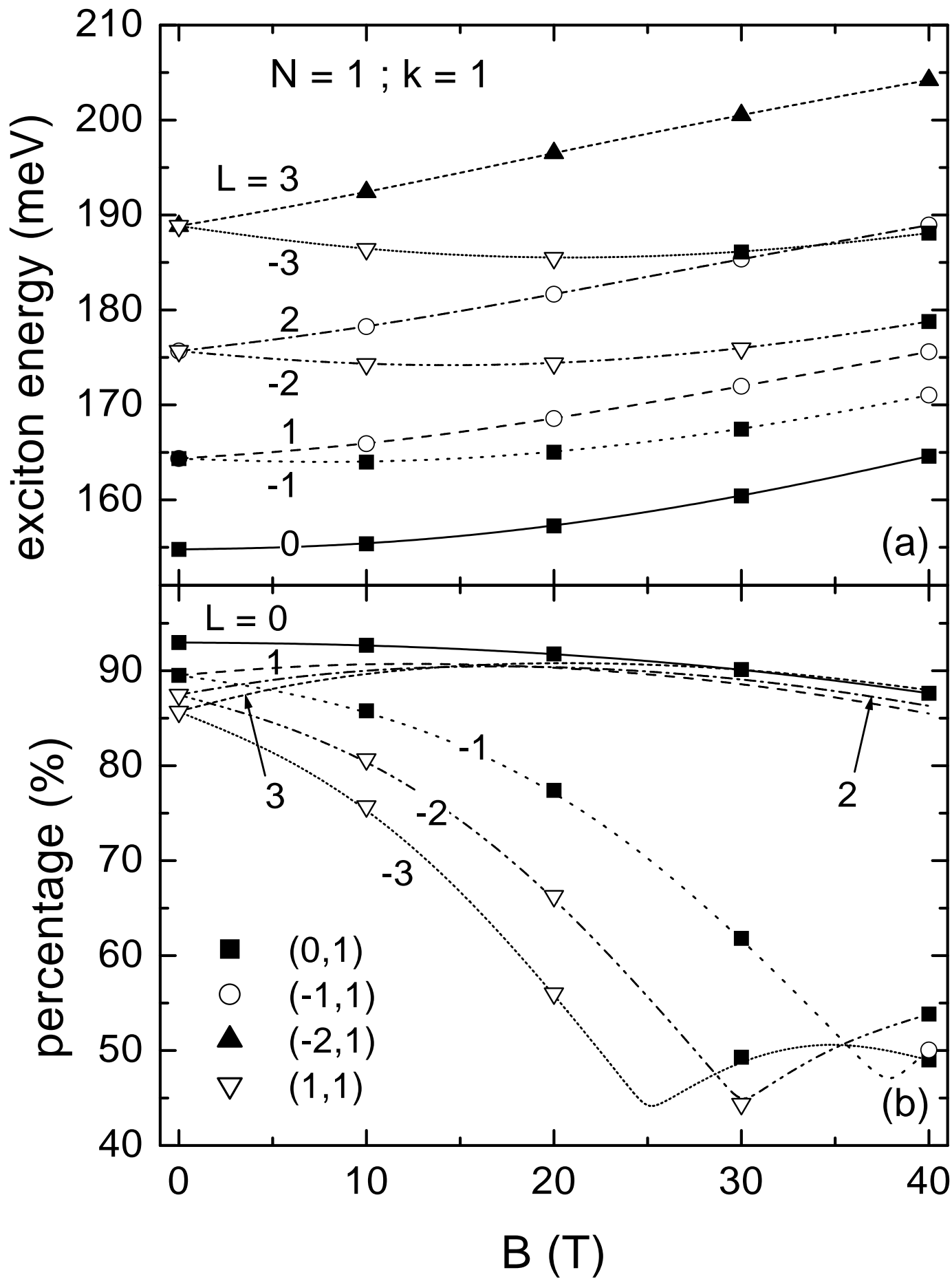
\begin{figure}[tbp]
\caption{(a) The energy states for different total angular momentum $L$ for
a disk with $R=8.95nm$ and $d=3.22nm.$ The symbols indicate the dominant
single-particle $(l,n)$ states and (b) gives the percentage of the
contribution of this term to the total wave function.}
\label{fig16}
\end{figure}

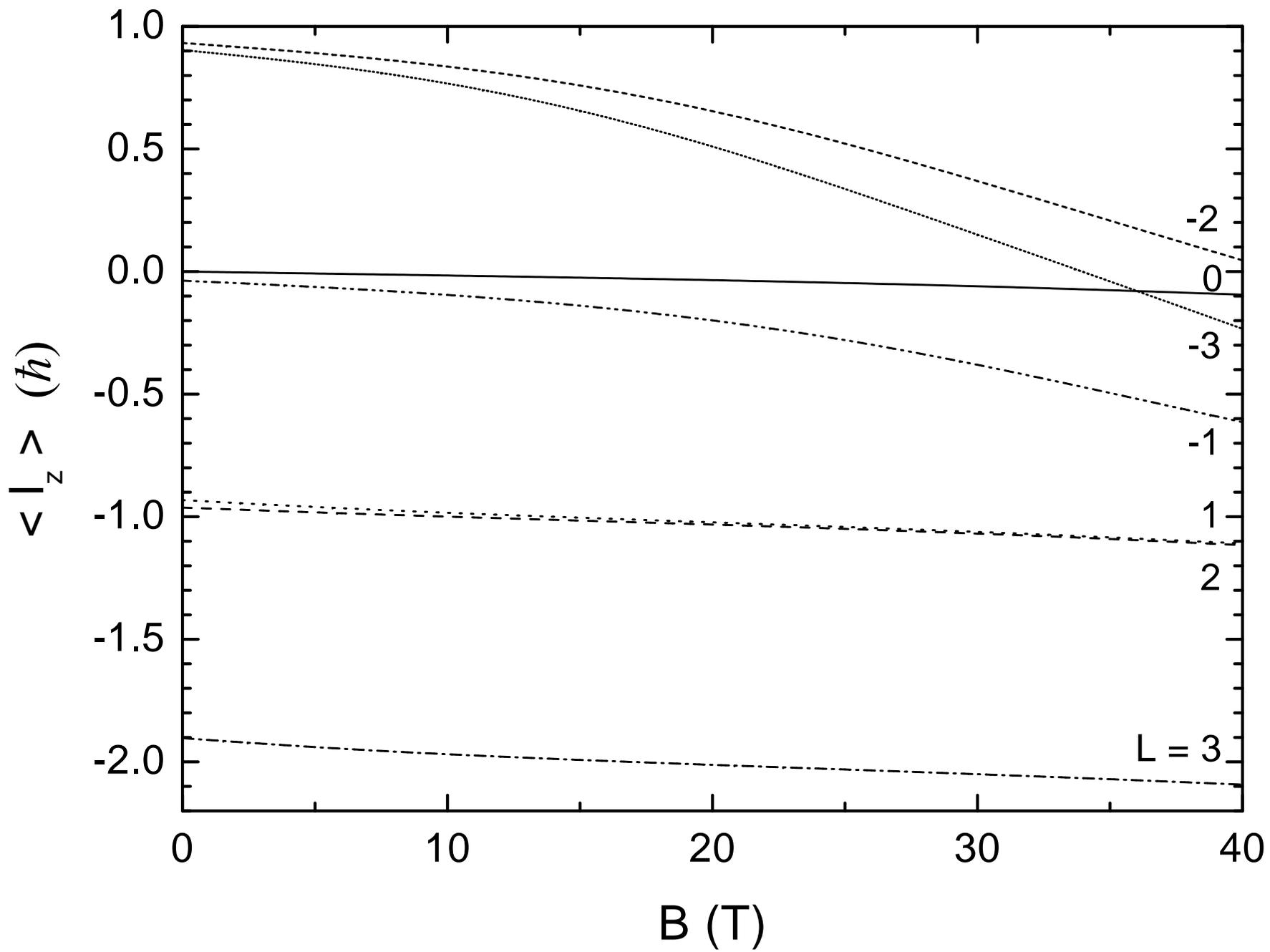
\begin{figure}[tbp]
\caption{The expectation value $\left\langle l_{z}\right\rangle $ for the
different $L$-states as a function of the magnetic fields.}
\label{fig17}
\end{figure}

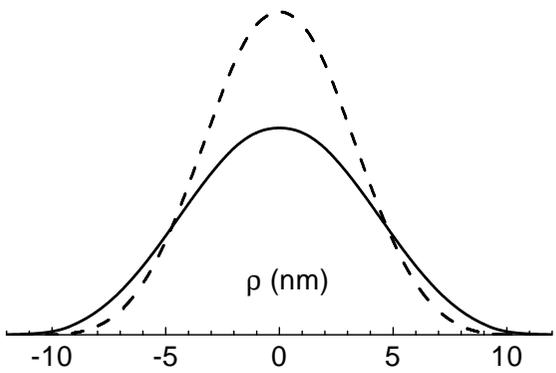
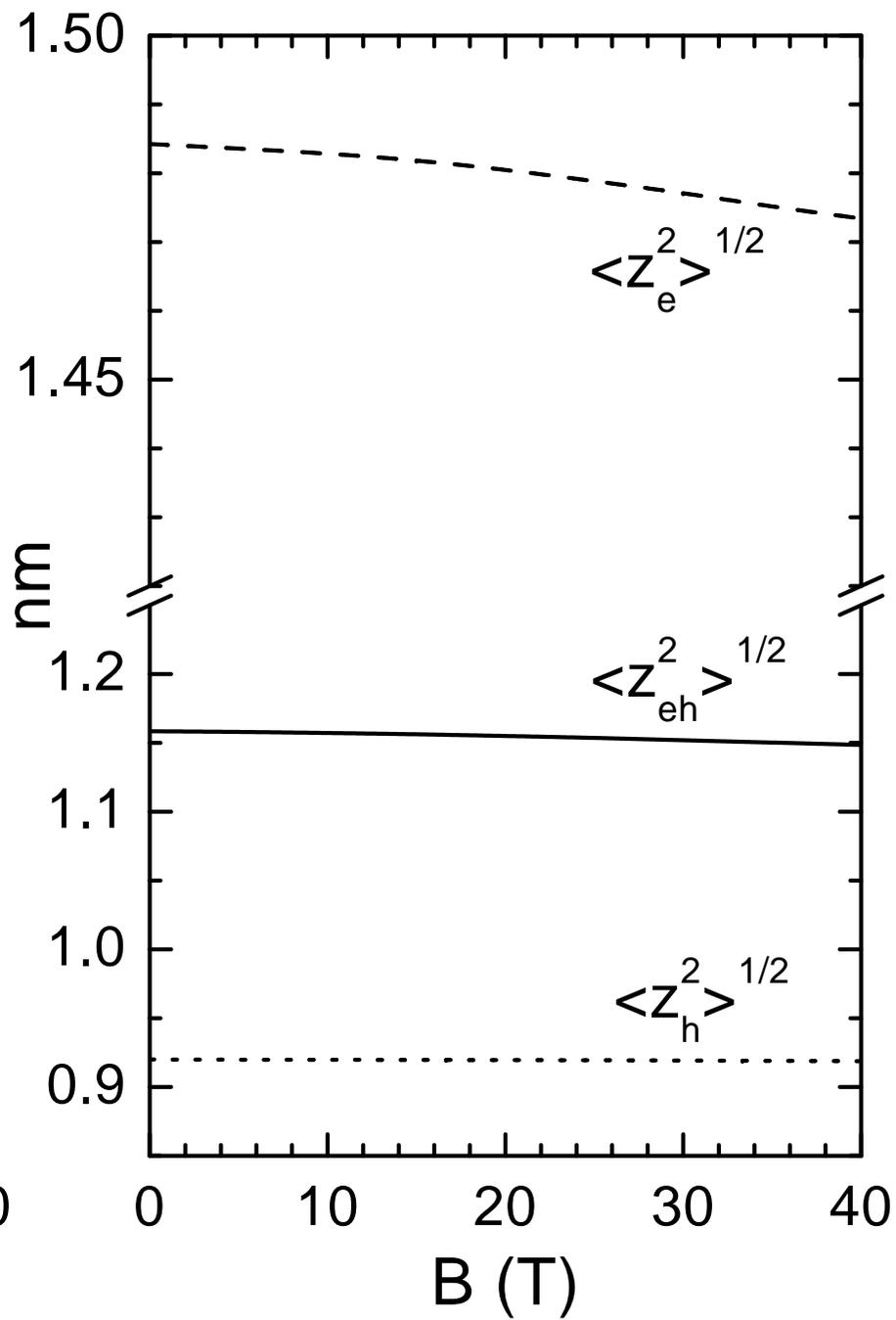
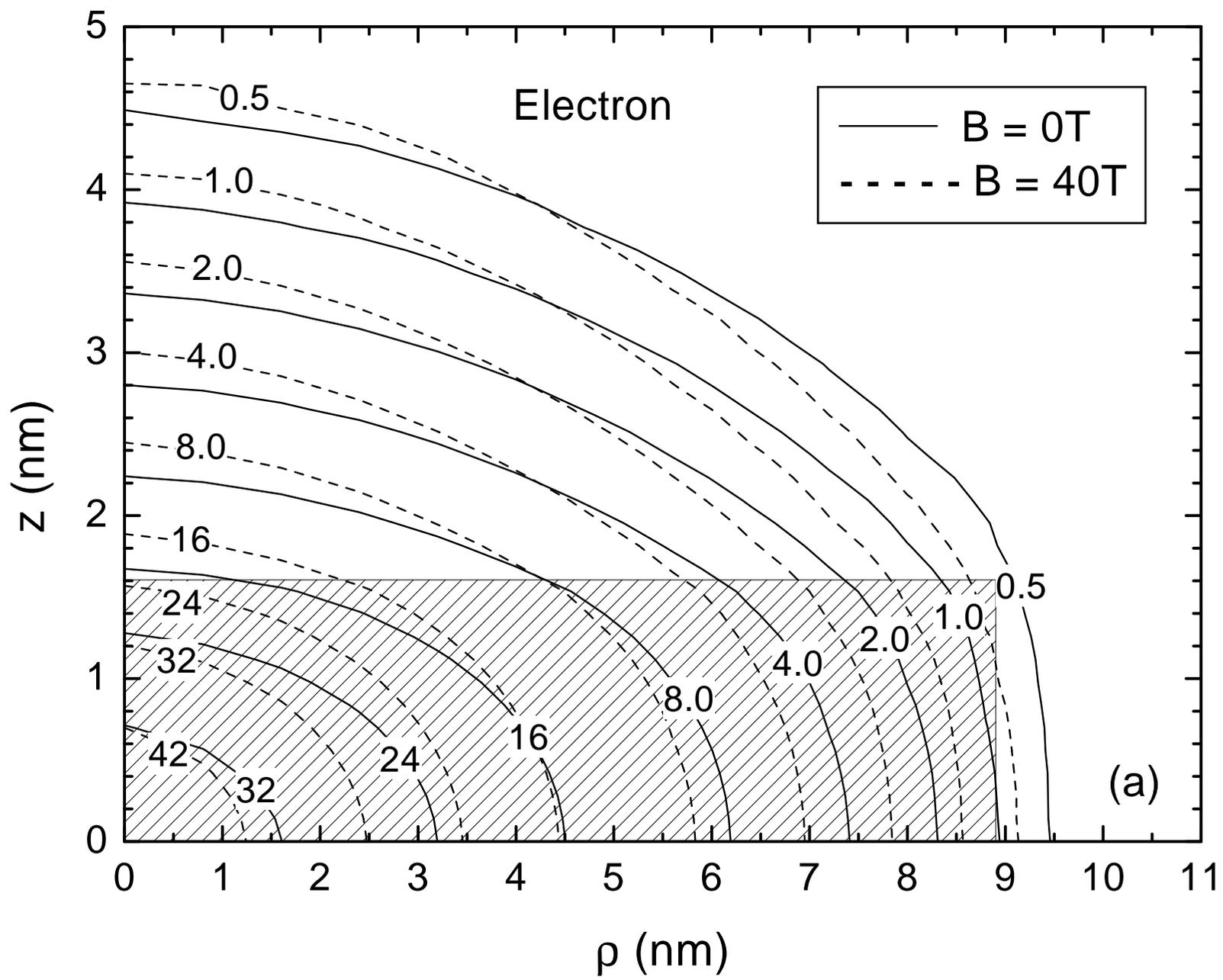
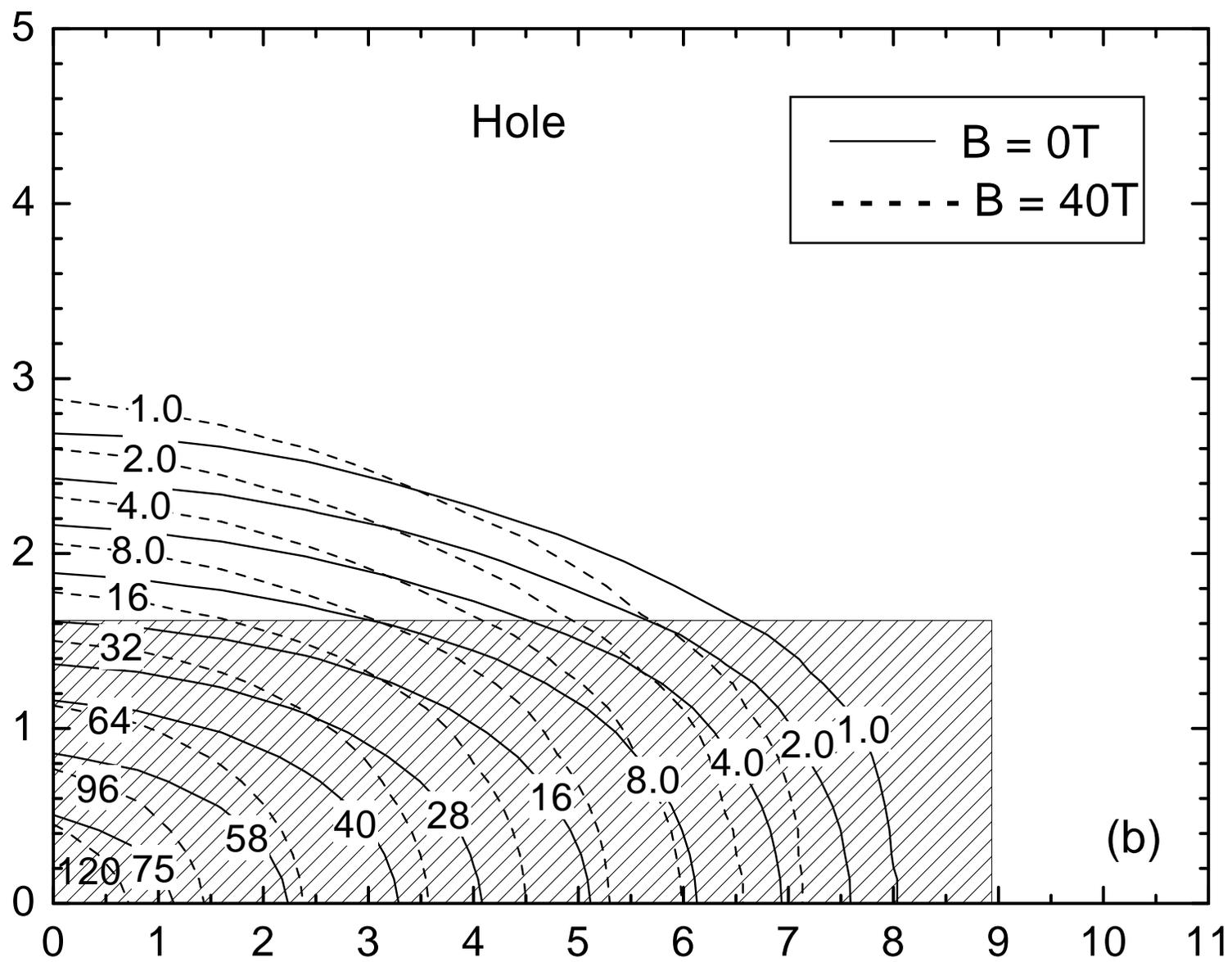
\end{document}